\documentclass[reprint,twocolumn,english,prl,notitlepage,nofootinbib,floatfix,longbibliography,superscriptaddress,aps]{revtex4-2}
\usepackage[T1]{fontenc}
\usepackage{siunitx}
\usepackage{comment}
\usepackage{graphicx}
\usepackage{placeins}
\usepackage{textcomp}
\usepackage{amssymb}
\usepackage{amsmath}
\usepackage{makecell}
\usepackage[dvipsnames]{xcolor}
\usepackage[english]{babel}
\usepackage{booktabs}
\usepackage[colorlinks=true,linkcolor=blue,urlcolor=blue,citecolor=blue,pdfusetitle]{hyperref}
\usepackage{tabularx} 
\usepackage{array} 
\usepackage{makecell} 
\usepackage{booktabs} 
\usepackage[normalem]{ulem}

\AtBeginDocument{\RenewCommandCopy\qty\SI}

\begin{document}

\title{Hybrid quantum network for sensing in the acoustic frequency range}

\author{Valeriy Novikov}
\thanks{These authors contributed equally}
\affiliation{Niels Bohr Institute, University of Copenhagen, Blegdamsvej 17, DK-2100 Copenhagen Ø, Denmark}
\affiliation{Russian Quantum Center, Skolkovo, Moscow, Russia}

\author{Jun Jia}
\thanks{These authors contributed equally}
\affiliation{Niels Bohr Institute, University of Copenhagen, Blegdamsvej 17, DK-2100 Copenhagen Ø, Denmark}

\author{Túlio Brito Brasil}
\thanks{These authors contributed equally}
\affiliation{Niels Bohr Institute, University of Copenhagen, Blegdamsvej 17, DK-2100 Copenhagen Ø, Denmark}

\author{Andrea Grimaldi}
\thanks{These authors contributed equally}
\affiliation{Niels Bohr Institute, University of Copenhagen, Blegdamsvej 17, DK-2100 Copenhagen Ø, Denmark}

\author{Maimouna Bocoum}
\affiliation{ESPCI Paris, PSL University, CNRS, Institut Langevin, Paris 75005, France}
\affiliation{Niels Bohr Institute, University of Copenhagen, Blegdamsvej 17, DK-2100 Copenhagen Ø, Denmark}

\author{Mikhail Balabas}
\affiliation{Niels Bohr Institute, University of Copenhagen, Blegdamsvej 17, DK-2100 Copenhagen Ø, Denmark}

\author{J\"{o}rg Helge M\"{u}ller}
\affiliation{Niels Bohr Institute, University of Copenhagen, Blegdamsvej 17, DK-2100 Copenhagen Ø, Denmark}

\author{Emil Zeuthen}
\affiliation{Niels Bohr Institute, University of Copenhagen, Blegdamsvej 17, DK-2100 Copenhagen Ø, Denmark}

\author{Eugene Simon Polzik}
\affiliation{Niels Bohr Institute, University of Copenhagen, Blegdamsvej 17, DK-2100 Copenhagen Ø, Denmark}

\begin{abstract}
Ultimate limits for sensing of fields and forces are set by the quantum noise of a sensor \cite{Tebbenjohanns2021,Purdy2013,Acernese2020}. Entanglement allows for suppression of such noise and for achieving sensitivity beyond standard quantum limits \cite{Yu2020,youssefi2023squeezed,Murch2008,Bohnet2014}. Applicability of quantum optical sensing is often restricted by fixed wavelengths of available photonic quantum sources. Another ubiquitous limitation is associated with challenges of achieving quantum-noise-limited sensitivity in the acoustic noise frequency range relevant for a number of applications. Here we demonstrate a novel tool for broadband quantum sensing by performing quantum state processing that can be applied to a wide range of the optical spectrum, and by suppressing quantum noise over an octave in the acoustic frequency range. An atomic spin ensemble is strongly coupled to one of the frequency-tunable beams of an Einstein-Podolsky-Rosen (EPR) source of light. The other EPR beam of light, entangled with the first one, is tuned to a disparate wavelength. Engineering the spin ensemble to act as a negative- or positive-mass oscillator we demonstrate frequency-dependent quantum noise reduction for measurements at the disparate wavelength. The tunability of the spin ensemble allows to target quantum noise in a variety of systems with dynamics ranging from kHz to MHz. As an example of the broadband quantum noise reduction in the acoustic frequency range, we analyse the applicability of our approach to gravitational wave detectors. Other possible applications include continuous-variable quantum repeaters and distributed quantum sensing.
\end{abstract}

\maketitle


Measurements on mechanical objects, spanning masses from pg to kg, and collective atomic systems, from ultracold atoms to room-temperature ensembles, have reached regimes dominated by quantum dynamics \cite{Tebbenjohanns2021,Purdy2013,Acernese2020,Yu2020,youssefi2023squeezed,Murch2008,Bohnet2014}. These systems are probed by light at optical or microwave wavelengths where the quantum-coherent interaction surpasses the thermal decoherence. The measurement imprecision due to the quantum fluctuations of light, and the measurement-driven disturbances known as the quantum backaction (QBA) collectively define the minimum total measurement noise called the Standard Quantum Limit (SQL). This constraint sets the limit on sensitivity of quantum sensors of motion \cite{Caves1980, Braginskii1996,  chen2013, Danilishin2019}, and restricts applications ranging from biomedical sciences \cite{Aslam2023} to the exploration of physics beyond the Standard Model \cite{Bass2024}.

Unlike a standard optomechanical oscillator \cite{markus}, an atomic spin oscillator can straightforwardly be engineered to exhibit an effective negative mass or, equivalently, frequency \cite{Julsgaard2001}. Theoretical proposals for measurements surpassing the SQL using an effective negative mass to counterbalance the QBA of a standard oscillator have been put forward \cite{Hammerer2009, Tsang2010, Tsang2012}. QBA evasion by coupling a spin oscillator to a macroscopic mechanical oscillator \cite{Moller2017, Thomas2021} and multi-tone readout of two micromechanical oscillators \cite{Lepinay2021}, enabling entanglement of the two systems, have been experimentally demonstrated. 
However, in those experiments both systems had to interact with light at a specific wavelength. When invoking the spin oscillator for QBA evasion, this wavelength is defined by the atomic transition. Another limitation of those initial demonstrations was the requirement for the oscillators' frequencies to be in the MHz range whereas the quantum noise reduction was achieved within a bandwidth much smaller than that frequency. Other methods of narrowband sub-SQL measurements include variational readout \cite{Kampel2017,Mason2019,Danilishin2019,Yu2020}.

Here, we report a hybrid quantum network for the broadband suppression of quantum noise beyond the SQL in the acoustic frequency range with a bandwidth of noise reduction that is comparable to the oscillation frequency. Our protocol relies on two key elements: a two-colour Einstein-Podolsky-Rosen (EPR) light source and a negative-mass spin ensemble. The EPR source establishes a quantum-entanglement link between the sensor and the negative-mass spin oscillator. The broad tunability of the spin permits us to tailor the frequency-dependence of the entanglement correlations and obtain the desired broadband quantum noise reduction. The system can, in principle, achieve complete quantum noise cancellation \cite{Khalili2018,Zeuthen2019}, while the sensing wavelength can be chosen across a broad range of the optical spectrum.
These features make the present approach useful in a number of applications where flexibility of the probe light wavelength and the sensor's eigenfrequency is required.

\begin{figure*}[ht!]
\centering
\includegraphics[width=0.99\textwidth]{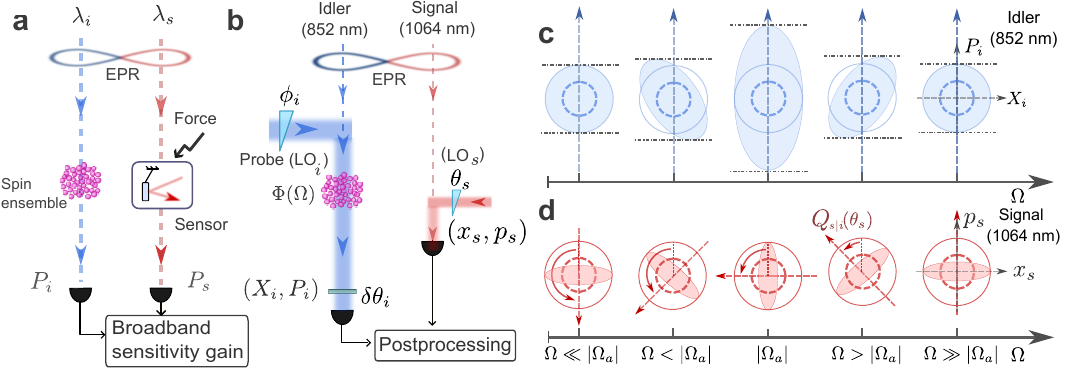}
\caption{\textbf{Parallel approach to broadband quantum noise suppression.} \textbf{(a)} General parallel scheme for the sensitivity enhancement of a quantum-noise-limited sensor. \textbf{(b)} Layout of the experimental setup.  The spin ensemble is probed by light at 852\,nm (idler) which is entangled with a 1064\,nm (signal) optical field. A suitably engineered atomic spin ensemble provides the required quadrature-rotation dynamics, described by $\Phi(\Omega)$. The two fields are analysed by the corresponding homodyne detectors with $\theta_{s}$ defining the quadrature phase of the signal field, and the phases $\phi_{i}$ and $\delta\theta_{i}$ defining the detection phase $\theta_{i}$ for the idler field. Conditioning the signal photocurrent on the idler which has interacted with the spin ensemble results in the noise reduction spectrum required for a particular application.  \textbf{(c)} The noise in the phase space for the idler field as a function of the frequency $\Omega$. Far away from $\Omega_a$ the idler noise (blue shaded area) corresponds to the EPR state. Closer to $\Omega_a$ the idler experiences a single-axis-twisting transformation due to interaction with the atomic spin ensemble. \textbf{(d)} Conditioning the signal on the idler by the post-processing of the photocurrents, as shown in panel b, results in conditional squeezing  in the signal observable $Q_{s|i}(\theta_{s})$ where the squeezing phase $\theta_{s}=\Phi(\Omega)$ changes with $\Omega$, as illustrated by the rotated red shaded ellipses in the lower panel. This process counteracts the quantum noise induced by optical probing of the force sensor by reducing the amplitude noise at $\Omega_a$ and reducing the phase noise far away from $\Omega_a$, thus enabling broadband quantum-enhanced sensing. Dashed circles indicate vacuum noise.}
\label{fig:1_motivation_experiment}
\end{figure*}

The main elements of the proposed quantum network are presented in Fig.~\ref{fig:1_motivation_experiment}a. We consider a sensor measuring a force $f$ and probed by light at wavelength $\lambda_s$, while an atomic oscillator with tunable dynamics interacts with light at wavelength $\lambda_i$. The light phase quadrature after passing through the sensor is $P_{s} = p_{s} + \mathcal{K}_s x_{s} + f$, where $(p_{s},x_{s})$, $(P_{s},X_{s})$ are input/output light canonical variables obeying the commutation relations $[x_{s}(\Omega),\, p_s(\Omega')]=[X_{s}(\Omega),\, P_s(\Omega')]=i\delta(\Omega+\Omega')$. 
In addition to the imprecision noise $\propto p_s$, the interaction induces the quantum backaction $\propto x_s$, which scales with the \emph{frequency-dependent} coupling strength $\mathcal{K}_s(\Omega)$. Similarly, after interaction with the atomic spin ensemble, the light phase quadrature reads  $P_{i} = p_{i} + \mathcal{K}_a(\Omega) x_{i}$. We demonstrate that we can engineer the spin oscillator to provide a counter-response of the sensor $\mathcal{K}_{a}(\Omega)=-\mathcal{K}_{s}(\Omega)$ corresponding to a negative-mass oscillator.  The ideal EPR state of light between signal and idler wavelengths $\lambda_s$ and $\lambda_i$ introduces correlations such that the uncertainties  $\Delta(p_{i} + p_{s}) \rightarrow 0$ and $\Delta(x_{i} - x_{s}) \rightarrow 0$. Thus, combining  the negative-mass response with EPR light it is possible to suppress both the quantum imprecision and backaction noise, $\Delta(P_{s} + P_{i}) = \Delta(p_{s}+p_{i}) + \mathcal{K}_s\Delta(x_{s}-x_{i}) + f \rightarrow f$, leading to broadband sensitivity gain.  

\section*{Hybrid quantum network}

\begin{figure}[ht!]
\includegraphics[width=0.48\textwidth]{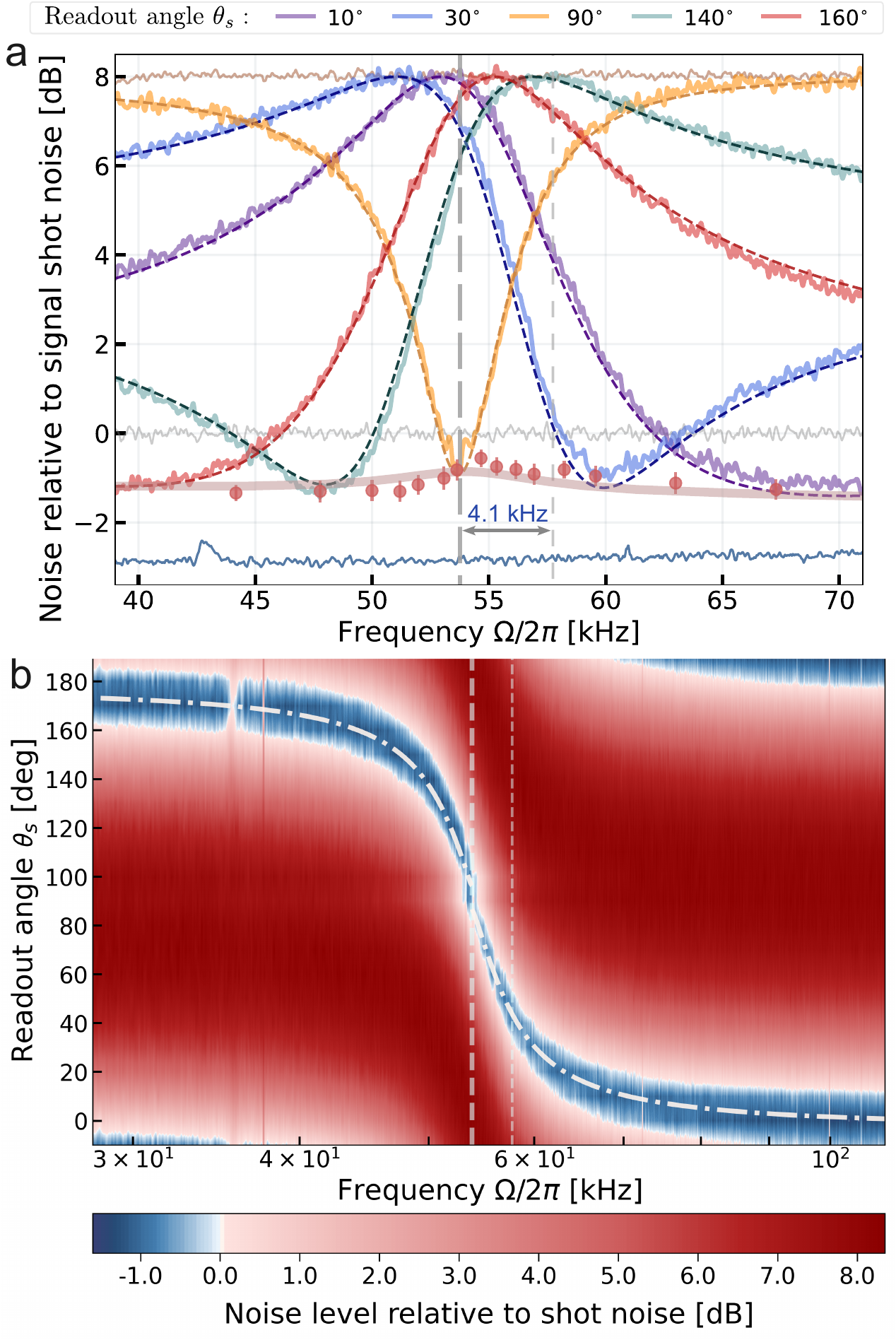}
\caption{\textbf{Experimental demonstration and theoretical fit of frequency-dependent conditional squeezing.} The spin ensemble is set to the positive-mass configuration with $\Omega_a/2\pi = 54$\,kHz (marked by the thick dashed vertical lines), and the idler field is detected at phase quadrature $\theta_i=0^{\circ}$. \textbf{(a)} Conditional noise relative to the signal shot noise for different readout angles $\theta_s$. The blue and grey traces represent frequency-independent conditional squeezing and signal-field shot noise, respectively, measured by setting the Larmor frequency to 1\,MHz. For each measurements phase, the dark dashed curves represent the fit with our theoretical model. The red data points with error bars are extracted from 60 samples around the minimal noise for each phase. The red solid curve is the conditional squeezing predicted by the model. \textbf{(b)} Contour spectrogram of the recorded conditional squeezing, illustrating its dependence on frequency $\Phi(\Omega)$, indicated by the white dashed-dotted curve. The bandwidth $\delta\Omega_\mathrm{SQL}/2\pi\approx4.1$\,kHz over which the squeezing phase rotates from $90^\circ$ to $45^\circ$ is indicated in both panels by the separation of the dashed vertical lines (see comments in the text).}
\label{fig:EntAtomsHybrid50kHzPosMass}
\end{figure}

The experimental implementation, depicted in Fig.~\ref{fig:1_motivation_experiment}b, involves generating an EPR state of two optical fields: an idler at 852\,nm and a signal at 1064\,nm. The idler passes through the spin oscillator prepared in an effective-negative-mass configuration. After the spins-idler interaction, we measure the quantum states of light in both the idler and signal arms. We demonstrate that applying the results of the homodyne measurement in the idler arm to the photocurrent in the signal arm, we obtain frequency-dependent conditional squeezing in the latter.  The optimal conditional squeezing is obtained at the readout phase angle $\Phi(\Omega)$ determined by the quantum dynamics of the spin oscillator, characterised by  $\mathcal{K}_a(\Omega)$. This conditioning is equivalent to the preparation of the squeezed state that compensates for the rotation of the signal output state induced by the target sensor's quantum backaction $\mathcal{K}_s(\Omega)$, thus maintaining noise reduction across all frequencies. 

While the approach we present is broadly applicable, we highlight its potential by examining quantum enhancements in Gravitational Wave Detectors (GWDs) \cite{Khalili2018}. Current GWDs are now operating near the SQL \cite{Yu2020, Acernese2020} and rely on frequency-dependent squeezing to broaden their observational range \cite{Ganapathy2023, Acernese2023}, and demonstrate squeezing of the quantum noise in the interferometer \cite{jia2024squeezing}. This technique \cite{Kimble2001}, which employs a 300-meter filter cavity to rotate the squeezed vacuum state, with up-to-10-km cavities planned in the future, is effective, but challenging to scale, particularly for the anticipated detuned dual-recycling Fabry-Pérot Michelson interferometer design of the next GWD generation \cite{Jones2020}. In this configuration, the detector susceptibility will require an additional hundred-meter-long filter cavity \cite{Danilishin2019}. Different approaches are currently under investigation \cite{Ma2017,Nishino2024,Brown2017}, but they are strongly limited by interdependencies with interferometer design \cite{Peng2024}. In this context, we propose a new method for generating frequency-dependent squeezing that achieves performance comparable to the filter cavity approach while offering a more compact and flexible setup. 

\section*{Atom-light two-colour quantum channel}

Our EPR source design, based on a Nondegenerate Optical Parametric Oscillator (NOPO) below threshold, enables the frequency-tunable entangled modes to couple to a specific target system (see Methods).  Here we choose the signal at $\lambda_{s} = 1064$\,nm  compatible with contemporary GWDs and the idler at $\lambda_{i}= 852$\,nm tuned to the $^{133}$Cs D$_2$ line. Due to the parametric amplification, the signal and idler will individually display a noise level higher than that of vacuum fluctuations (shot noise). EPR correlations allow the measurement on one mode to be used to infer the measurement outcome of another mode with precision below the vacuum fluctuations, which results in \emph{conditional squeezing} \cite{Reid1989}. For a given optical field $k$, the $x_k$ (amplitude) and $p_k$ (phase) quadratures define a general detection quadrature $q_k(\theta_k) = x_k \sin\theta_{k} + p_k \cos\theta_k$, selected by the homodyne angle $\theta_k$. We consider the case of measurements on the idler field used to estimate the outcome of measurements on the signal field. In this case, the signal field is conditionally squeezed if the inferred variance obeys $\text{Var}[q_{s}(\theta_s)+gq_i(\theta_i)] < 1/2$, thus manifesting the reduction of the noise compared to the vacuum level $1/2$.
The gain factor $g$ is selected so as to minimize the variance. In the output of the NOPO, correlations are maximized by the choice of quadratures satisfying the condition for detection angles $\theta_{i} = -\theta_{s}$ \cite{Reid1989}, where the phases are referenced to the pump beam phase (see Methods). For the signal and idler on resonance and the noise sideband frequency well within the NOPO cavity bandwidth, the conditional squeezing is \emph{frequency independent}.  The bottom blue trace in the top panel of Fig.~\ref{fig:EntAtomsHybrid50kHzPosMass} shows $3$\,dB of signal squeezing conditioned on the idler. For comparison with other experiments, we point out that we also measured $6$\,dB of two-mode entanglement according to the Duan-Simon criterion as shown in Extended Data Fig.~\ref{fig:Hybrid system calibration}a.  

Using optical pumping collinear with the magnetic field, we prepare the atomic ensemble in a highly polarized state, with the orientation of the collective atomic spin aligned along the bias field. The entangled idler beam is overlapped with an orthogonally polarized strong probe laser beam, whose polarization is adjusted to minimize the effect of higher-order light-atoms interactions \cite{Jia2023}. The strong beam also serves as the Local Oscillator (LO$_{i}$), see Fig.~\ref{fig:1_motivation_experiment}b and details in Extended Data Fig.~\ref{fig:setup_methods}. Both beams are detuned by 1.6\,GHz from the transition 6$^2$S$_{1/2}, F=4 \leftrightarrow $ 6$^2$P$_{3/2}, F'=5$.  The detuning provides sufficient quantum dispersive coupling to the atomic ensemble while significantly reducing the deleterious absorption. 
 
After interacting with the atomic ensemble, the quantum state in the idler field is measured with a polarimetric homodyne detector. The detection setup comprises a Quarter-Wave Plate (QWP) that introduces a phase offset $\delta\theta_{i}$ in the idler field [Fig.~\ref{fig:1_motivation_experiment}b]. This phase shift contributes to the detection phase $\theta_{i}$ and enables the observation of the ponderomotive squeezing induced by the atomic spin ensemble \cite{Jia2023}. The signal beam undergoes free propagation to the second homodyne detector, with the detected quadrature controlled by the relative LO$_{s}$ phase $\theta_s$. The photocurrents corresponding to each homodyne detector are recorded for postprocessing. A detailed description of the experimental setup is given in the Methods.

\section*{Tunable frequency-dependent conditional squeezing}

The atomic spin ensemble affects the phase quadrature of the idler beam in a frequency-dependent way 
\begin{equation}\label{eq:IO-P_i}
P_{i}(\Omega) = p_{i}(\Omega) + \mathcal{K}_a(\Omega) x_{i}(\Omega) + N_a(\Omega),
\end{equation}
while leaving the amplitude quadrature unchanged $X_i(\Omega) = x_i(\Omega)$ \cite{Moller2017}. $P_{i}(\Omega)$  contains contributions from both the input phase quadrature $p_i$ and the QBA which is proportional to $x_i$. In the context of spins, this transformation is called one-axis twisting \cite{borregaard2017one, kitagawa1993squeezed}, whereas in the context of light variables it is known as ponderomotive squeezing \cite{Danilishin2019}, which can be seen as an effective Kerr nonlinearity \cite{Bergman1991}. To achieve a broadband quantum noise reduction the one-axis twisting should be made frequency dependent.  Such frequency dependence is delivered by the atoms as  illustrated in Fig.~\ref{fig:1_motivation_experiment}c. Its frequency dependence is expressed by
\begin{equation}\label{eq:K_a}
    \mathcal{K}_a(\Omega)= \frac{\Gamma_{a}\Omega_{a}}{\Omega^{2}_{a} + (\gamma_{a}/2)^2 -\Omega^{2} -i \gamma^{}_{a}\Omega},
\end{equation} 
where the readout rate $\Gamma_a$ quantifies the interaction strength between the spin oscillator and light \cite{Jia2023}, $\gamma_{a}$ is the spin decay rate, and $\Omega_{a}$ is the Larmor frequency;  $N_a(\Omega)$ represents the atomic noise associated with the finite decay $\gamma_a>0$, contaminating the output state of light [Eq.~\eqref{eq:IO-P_i}]. The sign of $\Omega_a$ is determined by that of the effective oscillator mass, and can be switched from positive to negative by adjusting the bias field direction relative to the collective spin orientation, while varying the bias field strength tunes $|\Omega_a|$ \cite{Julsgaard2001, Moller2017}. 

Efficient broadband quantum noise reduction in our scheme requires strong light-spins coupling, which is characterized by high quantum cooperativity  $C_q=\Gamma_a/[\gamma_a(1+2n_{\text{th}})]$ \cite{Thomas2021}, where $n_{\text{th}}$ is the thermal occupation. We characterize the cooperativity by observing ponderomotive squeezing generated by the spin oscillator ranging from 1\,MHz down to 3\,kHz \cite{Jia2023} (see Methods and Extended Data Fig.~\ref{fig:Hybrid system calibration}b).

The effect of the spin oscillator on the quantum state of the idler light [Eq.~\eqref{eq:IO-P_i}] is, in our experiments, effectively a frequency-dependent rotation of the input quadratures determined solely by $\mathcal{K}_a(\Omega)$ [Eq.~\eqref{eq:K_a}]. 

We illustrate this for our experimental regime of slow spin decay $\gamma_{a}^2\ll\Omega_{a}^2,\Gamma_{a}^2$, in which it is permissible to let $\gamma_a\rightarrow 0$ in Eq.~\eqref{eq:K_a}; this yields a real $\mathcal{K}_a$ factor, equivalent to a purely in-phase atomic response (see Methods for details). 
Focusing on the quantum noise contribution of light in Eq.~\eqref{eq:IO-P_i}, we rewrite it as $P_i(\Omega)= z(\Omega)[p_{i}(\Omega)\cos\Phi(\Omega) - x_{i}(\Omega)\sin\Phi(\Omega)]$, thereby separating the atomic processing of light into a quadrature rotation
\begin{equation}\label{eq:Phi}
\Phi(\Omega)=-\arctan\{\mathcal{K}_a(\Omega)\},
\end{equation}
and an overall amplification factor $z(\Omega)=\sqrt{1+\mathcal{K}^2_a(\Omega)}$. 
We obtain the signal-arm measurement photocurrent \emph{conditioned} on the measurements of the idler quadrature $P_i$ as
\begin{equation}
    Q_{s|i}(\theta_{s},\Omega) = q_{s}(\theta_{s},\Omega) + g(\Omega)P_i(\Omega).
\end{equation}
Using a suitably optimized gain $g(\Omega)$, we achieve \emph{conditional} squeezing of the idler beam with \emph{frequency-dependent} squeezing angle $\Phi(\Omega)$ as shown in Fig.~\ref{fig:EntAtomsHybrid50kHzPosMass}b, corresponding to the rotation induced on the idler by the spin ensemble [Fig.~\ref{fig:1_motivation_experiment}c]. That is, for each frequency component $\Omega$ there exists an optimal signal detection phase $\theta_{s}=\Phi(\Omega)$ required for witnessing the optimal noise reduction.

The range of $\Phi(\Omega)$ spans from $0^{\circ}$ to $180^{\circ}$ as the Fourier frequency is scanned from $\Omega\gg|\Omega_{a}|+\Gamma_{a}$ to $\Omega\ll|\Omega_{a}|-\Gamma_{a}$, crossing $\theta_{s} = 90^{\circ}$ at the Larmor frequency $\Omega =|\Omega_{a}|$. Thus, the squeezing ellipse undergoes complete rotation over the frequency range with the bandwidth $2\Gamma_{a}$. Hence, the dependence of the rotation on the frequency can be changed by modification of the Larmor frequency $\Omega_{a}$ and spin readout rate $\Gamma_{a}$. The atomic noise $N_a(\Omega)$ in Eq.~\eqref{eq:IO-P_i} does not influence the squeezing angle $\Phi(\Omega)$, but merely degrades the degree of conditional squeezing (see Methods).

In Fig.~\ref{fig:EntAtomsHybrid50kHzPosMass} we show the noise reduction obtained with the spin oscillator in the effective positive-mass configuration; the observed behaviour is consistent with the squeezing-angle dependence $\Phi(\Omega)$ required for quantum noise reduction in a negative-mass sensor.

Reversing the sign of the effective mass changes the sign of $\Omega_{a}$ in $\mathcal{K}_a$ [Eq.~\eqref{eq:K_a}] and consequently, the direction of the frequency-dependent rotation $\Phi(\Omega)$ of the squeezing [Eq.~\eqref{eq:Phi}]. Specifically, the effective negative mass of the spin oscillator opens up the opportunity to counterbalance the quantum backaction from radiation pressure in quantum optomechanics. In the limit $|\Omega^{}_{a}$|, $\gamma^{}_{a}\ll\Omega$, the phase $\Phi(\Omega)=-\arctan{(\Gamma^{}_{a}\Omega^{}_{a}/\Omega^{2})}$ mimics the optomechanical interaction with a free-mass object \cite{Danilishin2019}. The conditional squeezed state thus becomes compatible with GWDs in a simple Michelson configuration and potentially allows for broadband quantum noise reduction in such an interferometer. 

\begin{figure*}[ht!]
\centering
\includegraphics[width=0.99\textwidth]{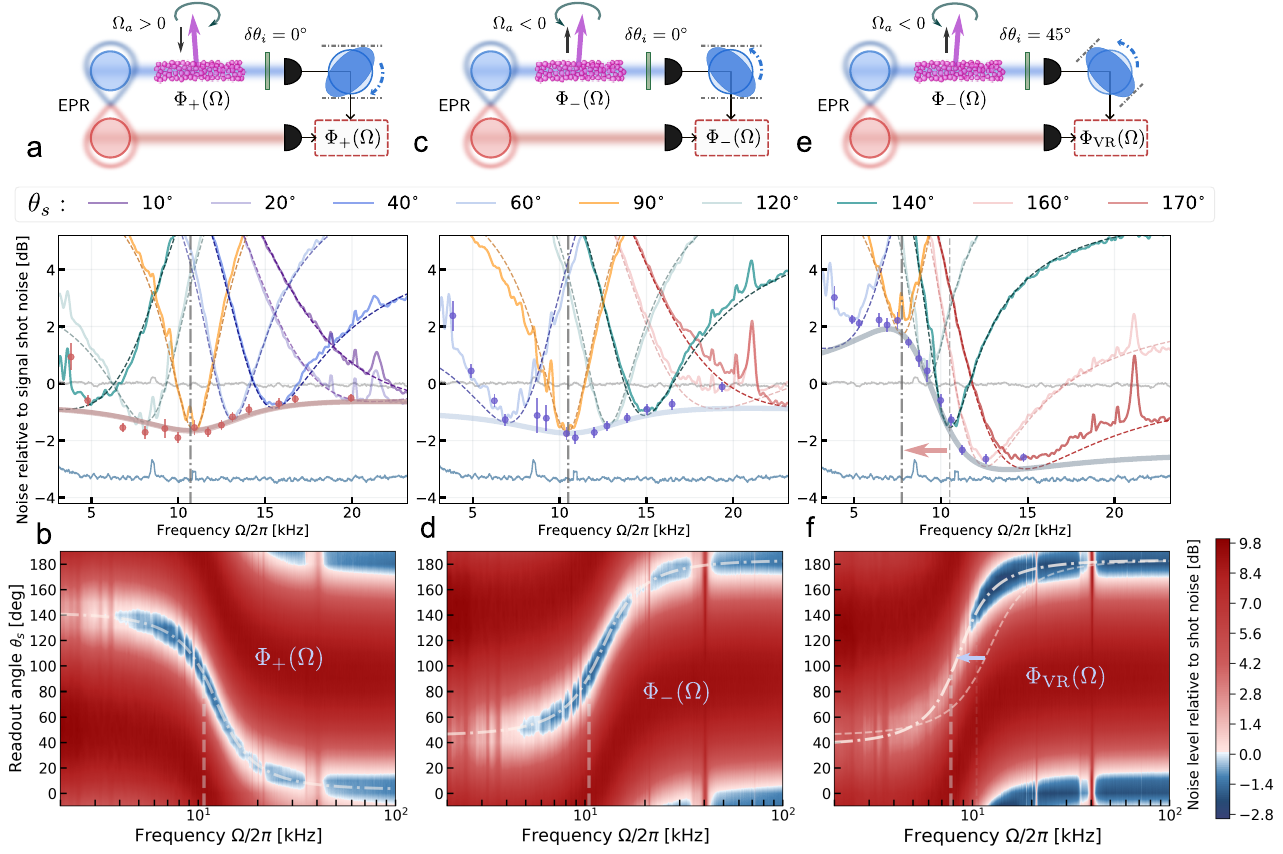}
\caption{\textbf{Frequency-dependent conditional squeezing at acoustic frequencies.} The inferred quantum noise for the signal field in three spin-ensemble configurations in units of the signal-field shot noise. Top plots (a,c,e) show inferred noise as a function of Fourier frequency, recorded at different readout angles $\theta_s$ for specific ensemble configurations. In each plot, the data points represent the measured minimal conditional noise for each readout angle, and the associated thick solid curve shows the prediction from our model. The blue traces at $-3$\,dB in the upper plots display frequency-independent conditional squeezing, measured by detuning the Larmor frequency $|\Omega_a|/2\pi$ up to 1\,MHz.  By combining the data from the joint measurements over the continuous phase range ($0^{\circ}$ to 180$^{\circ}$), we construct the bottom contour spectrograms. \textbf{(a, b) Positive mass:} The frequency dependence of squeezing angle $\Phi_{+}(\Omega)$ with $\Omega_a/2\pi = 10.7$\,kHz and the phase offset $\delta\theta_{i}=0^{\circ}$ adjusted by quarter-wave plate in the idler optical mode. \textbf{(c, d) Negative mass:} $\Omega_a/2\pi=-10.5$\,kHz and $\delta\theta_{i}=0^{\circ}$, showing opposite frequency-dependence $\Phi_{-}(\Omega)$ to the positive mass. \textbf{(e, f) Negative mass with virtual frequency shift:} for the same ensemble parameters as in (c, d), the phase offset is set to $\delta\theta_{i}\approx45^{\circ}$. The gray dash-dotted vertical line, representing the frequency corresponding to the minimum quantum noise for the signal phase of $\theta_{s}=90 ^{\circ}$ in (e), indicates an effective frequency downshift of $\approx 2.7$\,kHz (as shown by red arrow) relative to the Larmor frequency, marked by the dashed line. In (f), this frequency downshift also sharpens the frequency-dependent phase rotation [as indicated by the dashed-dotted curve $\Phi_{\text{VR}}(\Omega)$ relative to the dashed curve $\Phi_{-}(\Omega)$] . Meanwhile, we observe enhanced squeezing levels above the Larmor frequency $|\Omega_{a}|$ compared to (c,d). The total detection phase $\theta_{i}=0^\circ$ was chosen for each of three configurations outlined above. See Extended Figs. 4,5 for the analysis of the error bars. 
}
\label{fig:EntAtomsHybrid10kHzPosNegMasVR}
\end{figure*}

\section*{Quantum noise suppression in the acoustic frequency band}

The optimal gain $g(\Omega)$ is applied for each $\theta_{s}$ to maximise the noise suppression based on the entire idler measurement record (see Methods). The noise reduction below the signal beam shot noise level in a broad frequency range confirms that the atomic ensemble imposes the phase shift on the conditional squeezing angle ranging from $\theta^{}_{s}=0^{\circ}$ to $\theta^{}_{s}\approx$ $180^{\circ}$. As shown in Fig.~\ref{fig:EntAtomsHybrid50kHzPosMass}, the theoretical noise spectra calculated based on independent calibrations as described in Methods, coincide well with the experimental data. The quantum noise suppression does not reach the level set by the initial entangled state of light due to the broadband spin noise (see Methods) at higher Fourier frequencies and due to the atomic thermal noise near $\Omega^{}_{a}$.

Following the approach used in Fig.~\ref{fig:EntAtomsHybrid50kHzPosMass}a, 18 conditional squeezing measurements are combined to construct a contour spectrogram, where the conditional noise is presented as a function of both signal homodyne phase $\theta_{s}$ and Fourier frequency $\Omega$. The result in Fig.~\ref{fig:EntAtomsHybrid50kHzPosMass}b shows the mapping of the frequency-dependent rotation $\Phi(\Omega)$. The observed trajectory matches well with the theoretical model (white dashed-dotted curve) based on the calibrated experimental values of $\Gamma_{a}$,  $\gamma_{a}$ and $\Omega_{a}$ (see Methods). The model predicts that in the limit of $|\Omega_{a}|\gg\Gamma_{a},\gamma_{a}$, the bandwidth $\delta\Omega_\mathrm{SQL}>0$ over which the squeezing phase rotates from the amplitude quadrature [$\Phi(\Omega_a)=90^\circ$] to halfway between amplitude and phase quadratures [$\Phi(\Omega_a+\delta\Omega_\mathrm{SQL})=90^\circ\mp\mathrm{sign}(\Omega_a)45^\circ$] is $\delta\Omega_{\text{SQL}}/2\pi\approx \Gamma_{a}/4\pi = 4.1$\,kHz. As shown in Fig.~\ref{fig:EntAtomsHybrid50kHzPosMass}a, this value aligns well with the experimentally observed frequency difference between the Larmor frequency and the Fourier frequency, at which the squeezing ellipse is rotated by 45$^{\circ}$. 

After validating the model, we test the system tunability by lowering the oscillator frequency into the audio band,  $|\Omega_{a}|/2\pi \approx 10$\,kHz. We optimised the system, increasing the stability of the EPR source and its coupling to the atomic spin oscillator. The results are shown in Fig.~\ref{fig:EntAtomsHybrid10kHzPosNegMasVR}.  For the positive-mass spin configuration (a,b), we confirm the broadband conditional squeezing capabilities of our system by measuring the bandwidth of the phase shift $\delta\Omega_{\text{SQL}}/2\pi \approx 4.1$\,kHz comparable to the central frequency $|\Omega_{a}|$ and in agreement with the theoretical model (see Methods). We then optically pump the spin oscillator into the negative-mass configuration (c,d). Here, we demonstrate the ability to invert the direction of the phase shift of conditional squeezing while keeping all the other reference parameters unchanged. The results displayed in Fig.~\ref{fig:EntAtomsHybrid10kHzPosNegMasVR}d correspond to the desirable configuration for quantum noise reduction in optomechanical sensors including GWDs, since the produced rotation of the squeezing phase $\Phi_{-}(\Omega)$ counteracts the QBA effect for a positive-mass quantum system, such as an interferometer mirror. 

The phase offset $\delta\theta_{i}$ in the entangled idler  field due to the QWP, shown in the upper panel of Fig.~\ref{fig:EntAtomsHybrid10kHzPosNegMasVR}e, impacts the functional dependence of $\Phi(\Omega)$. The implications of rotating the quadrature basis after probing the spin ensemble are shown in the right column of Fig.~\ref{fig:EntAtomsHybrid10kHzPosNegMasVR}. We highlight the decrease of the effective spin oscillator frequency by $\approx2.7$\,kHz and the change of the bandwidth $\delta\Omega_{\text{SQL}}$ of the squeezing rotation down to $\approx2.5$\,kHz, comparing the spectrogram in Fig.~\ref{fig:EntAtomsHybrid10kHzPosNegMasVR}f for $\delta\theta_i\approx 45^{\circ}$ to Fig.~\ref{fig:EntAtomsHybrid10kHzPosNegMasVR}d where $\delta\theta_i=0^{\circ}$ was set. The observed phenomenon is attributed to the \emph{virtual rigidity} effect \cite{Danilishin2012}, which effectively modifies the oscillator's response to the quantum noise of the probe light as if its resonance frequency and the readout rate were changed \cite{Jia2023}, leading to $\Phi(\Omega) \rightarrow \Phi_{\text{VR}}(\Omega)$. This effect, rooted in the atomic ponderomotive squeezing and an instance of variational measurement, also alters the degree of conditional quantum noise reduction (blue data points) for different parts of the spectrum, as displayed in Fig.~\ref{fig:EntAtomsHybrid10kHzPosNegMasVR}c (for details see Supplementary Material Sec.~I).
Overall, the control of the phase shift $\delta\theta_{i}$ in the entangled idler field provides additional flexibility for tuning the features of frequency-dependent squeezing and for corresponding optimization of the quantum noise reduction \cite{Zeuthen2019}.

The approximation $\mathcal{K}_a\approx\Gamma^{}_{a}/\Omega^{}_{a}$ for the atomic coupling factor applies at $\Omega\ll|\Omega_{a}|$. Hence, as the Larmor frequency approaches the readout rate, the full phase rotation range for the conditional squeezing cannot be achieved for the upper (lower) range with the positive (negative) mass spin oscillator. This limitation on the phase rotation $\left[0^{\circ}, 140^{\circ} \right]$ is seen in Fig.~\ref{fig:EntAtomsHybrid10kHzPosNegMasVR}(b,d), despite the increased technical noise near DC frequencies. However, the readout rate and the decay can be adjusted by carefully engineering the light-spins interaction \cite{Jia2023}.

\section*{Outlook}

We have demonstrated that a tunable entangled light source coupled to a spin oscillator with an adjustable oscillation frequency allows, in principle, for the reduction of both the shot noise and QBA noise of the measurement. This reduction can be achieved at widely tunable optical wavelengths and within a broad noise frequency range. These features make our hybrid quantum system a promising candidate for the sensitivity improvement in the next generation of GWDs. 

Although our current results exhibit less conditional squeezing compared to the original frequency-independent squeezing level, our model suggests that reducing the broadband spin noise by a factor of six and the spin thermal occupation by a factor of three would be sufficient to achieve the frequency-dependent squeezing level across the full bandwidth limited only by the degree of EPR correlations of light as illustrated by Extended Data Fig.~\ref{fig:conditional squeezing with improved system}. The reduction in thermal occupation is a combination of optical pumping improvement and laser probe classical noise mitigation \cite{Jia2023} (see details in Supplementary Material Sec.~II~B). The degree of observed EPR correlations can be further increased by reducing the overall optical losses and mitigating the optical phase fluctuations, allowing higher  parametric gain in the NOPO \cite{Brasil2022}.

With the atomic spin oscillator maintaining quantum-noise-limited performance down to the GW backaction-dominant regime, our system is analogous to the filter cavity implementations \cite{Ganapathy2023}. We infer that the squeezing phase rotation provided by the 8-cm-long cell in the present work is equivalent to the rotation imposed by a 5-m-long Fabry-Pérot filter cavity with finesse $\sim$ 6000 \cite{Lee2023}, as follows from the results shown in Fig.~\ref{fig:EntAtomsHybrid10kHzPosNegMasVR}d [see Supplementary Material Sec.~III].  The length of the equivalent optical cavity is effectively extended to 10\,m when the virtual rigidity downshift in Fig.~\ref{fig:EntAtomsHybrid10kHzPosNegMasVR}f is applied. 

The steps required for application of our method in the bandwidth of the existing GWDs mainly involve reduction of the acoustic noise. It can be achieved, for example, by placing the experimental setup in vacuum, similar to the filter cavities of LIGO and Virgo. Under such conditions, the lower spectral bound of the QBA-dominated performance of our system can be extended to 1--2\,kHz. Using the 1--2\,kHz shift of the response function to lower frequencies by virtual rigidity demonstrated in the present paper, the effective quantum noise reduction down to the low-frequency end of LIGO and Virgo can be achieved \cite{Zeuthen2019}. Another condition for improved low-frequency performance of our system --- the long coherence time of the spin oscillator –-- has been already achieved at our lab using the atomic cell with the cross section of 5$\times$5\,mm$^2$ for which 6\,Hz intrinsic decoherence rate has been observed \cite{Jia2023}.

The spin system is flexible in both its resonance frequency and bandwidth, and is scalable due to its relative compactness and simplicity. The protocol can be extended using multiple ensembles and in this way more intricate quantum noise reduction spectral profiles $\Phi(\Omega)$ within a required spectral range can be engineered. In the context of the quantum noise reduction in GWDs, a cascade of spin ensembles resembles the implementation of a cascade of filter cavities \cite{signalrecycling, Danilishin2019}. 

The EPR source design can be adapted to link our spin oscillator with systems ranging from optomechanics to atomic physics. The required signal wavelength ranging from 700\,nm to 2000\,nm can be achieved by using a suitable signal laser and a corresponding nonlinear medium for parametric downconversion. Leveraging this opportunity, we envision potential for other applications in the field of quantum-enhanced metrology, such as detection of quantum motion of nanoparticles and cantilevers \cite{recoilheat, yap2020broadband}, where broadband, frequency-dependent engineering of the quantum noise of light is required. 

To the best of our knowledge, the demonstrated system is the first combination of multicolour continuous-variable (CV) entangled light modes with a quantum memory \cite{Julsgaard2004,Jensen2010}, which is the backbone of proposals for a CV quantum repeater \cite{repeater}. The versatility of our hybrid network enables quantum protocols coupling efficient atom-light interfaces \cite{hammerer2010quantum} to systems ranging from the nano to the macro scale \cite{Kimble2008,Wehner2018}. 




\appendix

\section*{Methods}

\setcounter{figure}{0}
\renewcommand\theHfigure{sec.\thefigure}
\renewcommand{\figurename}{Extended Data Fig.}
\renewcommand{\tablename}{Extended Data Table}

A detailed layout of the experimental setup is shown in Extended Data Fig.~\ref{fig:setup_methods}. Below we discuss the main components of the system and their characterization.

\subsection*{EPR source}

As shown in Extended Data Fig.~\ref{fig:setup_methods}, two lasers at 852\,nm (Msquared - Solstis) and 1064\,nm (Innolight - Mephisto) drive the Sum-Frequency-Generation (SFG) producing the pump beam at 473\,nm for a Nondegenerate Optical Parametric Oscillator (NOPO) which generates the EPR state of light. The NOPO cavity design and the operation principle are reported in Ref.~\cite{Brasil2022}. The cavity has a bow-tie configuration with a Periodically Poled Potassium Titanyl Phosphate (PPKTP) nonlinear crystal. The resonance of the NOPO for the 852\,nm and 1064\,nm downconverted modes is maintained by locking the cavity to the 1064\,nm laser and locking the 852\,nm laser to the cavity, using beams counter-propagating to the EPR modes. To achieve the quantum noise limited performance in the audio frequency band required in the present paper, we have implemented the reduction of classical noise contamination on our probe laser using an active noise eater and have achieved robust control of the EPR state phases, as described in depth in Ref.~\cite{UpComingEPR2}. 
 
The EPR output modes of the NOPO cavity are separated by a dichroic mirror. The signal beam at 1064\,nm is mixed with the corresponding LO$_s$ on a 50:50 beam splitter, and the canonical operators $(x_{s}, p_{s})$ are measured using a balanced homodyne detector HDs (Extended Data Fig.~\ref{fig:setup_methods}). The 852\,nm idler beam is combined with an orthogonally polarized probe beam LO$_i$ on a polarizing beam splitter (PBS) and polarization homodyning is performed.  

We observe $9$\,dB of two-mode squeezing, corresponding to $6$\,dB of conditional squeezing of the signal field when the two EPR beams are analysed directly (the idler bypasses the atoms) \cite{UpComingEPR1}. The electronic noise floor is more than $17$\,dB below the shot noise level.

Prior to entering the atomic ensemble, the combined idler beam and LO$_i$ are shaped into a square top-hat profile, enabling the optimal readout of atomic spins (see the \emph{Atomic spin oscillator} section below). To characterize the overall propagation losses, including those from the cell windows, beam shaper and optics, a measurement with the Larmor frequency tuned to 1\,MHz, beyond the detection frequency band, has been performed. We still observe $\sim 3.3$\,dB conditional squeezing from 3\,kHz to 60\,kHz relative to the signal vacuum noise, as shown in the upper panel (a) of Extended Data Fig.~\ref{fig:Hybrid system calibration}. Noise peaks at 26\,kHz and 36\,kHz are experimental artefacts caused by the intensity noise eater. The theory of the conditional squeezing, detailed in the Supplementary Material Sec.~I, allows us to extract a squeezing factor $r=1.42$ and unbalanced detection efficiency of $\eta_{s} \sim 0.92$ for the signal arm and $\eta_{i} = \eta_{i,\mathrm{out}}\eta_{i,\mathrm{in}} \sim 0.8$ for the idler arm by fitting the measured power spectrum densities. See Extended Data Table~\ref{tab:hybrid quantum network} for parameter values.

\subsection*{Atomic spin oscillator}

The atomic spin oscillator is implemented by confining a gas of $N_{\text{Cs}}\sim 10^{10} $ $^{133}$Cs  atoms in a $2 \times 2 \times 80$\,mm$^3$ glass channel inside a vapour cell. The cell windows are anti-reflection coated and the inner walls are coated with an anti-relaxation paraffin material to minimize the decoherence from spin-wall collisions.  The atomic vapour density in the glass cell is defined by the temperature of the caesium droplet in the stem of the cell, which was set to  $\sim40^{\circ}$C in the experiment.
A multilayer magnetic shield around the cell provides isolation from the Earth's magnetic field and other high-frequency magnetic noise, while a set of coils running low-noise DC currents generates a highly homogeneous bias magnetic field within the cell. By adjusting the DC current, we can control the Larmor frequency $|\Omega_a|$ from $1$\,MHz down to $3$\,kHz, maintaining quantum-dominated performance throughout this frequency spectrum \cite{Jia2023}.  

\begin{figure*}[htp]
\centering
\includegraphics[width = 0.99\textwidth]{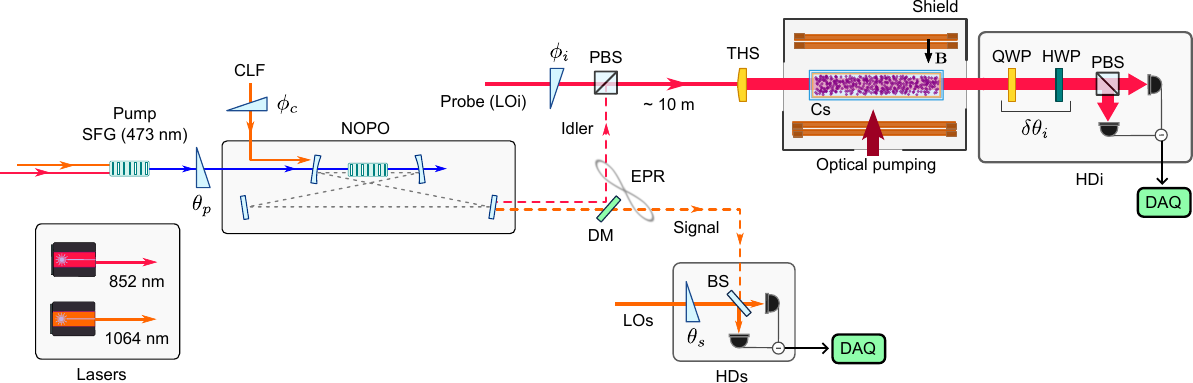}
\caption{\textbf{Experimental Setup Main Components.} The experimental setup includes two lasers: a 1064\,nm continuous-wave (CW) laser and a tunable CW Ti:Sapphire laser at 852\,nm. The two-colour Einstein-Podolsky-Rosen (EPR) state setup involves two nonlinear optical interactions: sum-frequency generation (SFG) and a non-degenerate optical parametric oscillator (NOPO). The 1064\,nm and the 852\,nm laser produce the SFG light in a nonlinear crystal, which serves as the pump for the NOPO with pump phase $\theta_p$. The NOPO generates the EPR state at 1064\,nm (signal) and 852\,nm (idler), represented by dashed lines. The signal and idler beams are separated by a dichroic mirror. The signal is directed to balanced homodyne detection (HDs) with local oscillator phase $\theta_s$, while the idler is mixed with the probe beam at a polarizing beam splitter (PBS) with relative phase $\phi_i$ and propagates in free space ($\sim 10$\,m) to the atomic oscillator setup. The probe-idler spatial profile is modified to a square top-hat beam by a top-hat shaper (THS) before being sent to the caesium (Cs) vapour cell. The cell is placed inside a magnetic shield with a set of coils controlling the bias magnetic field. The optical pump system prepares the spin ensemble in a highly polarized state. After interaction, the idler quantum state is sent to polarization homodyne detection (HDi) where a quarter-wave plate (QWP) and half-wave plate (HWP) determine the phase shift $\delta\theta_i$. A portion of the 1064\,nm laser is frequency-shifted by 3\,MHz to produce the coherent-lock field (CLF) beam injected to the NOPO with phase $\phi_c$, which provides the phase reference for the detection system and feedback control of the detected quadratures. The fields for the local oscillators LOs and LOi are produced by the 1064\,nm and 852\,nm lasers, respectively. Each LO is filtered by a mode-cleaner cavity (not shown). Both photocurrents from the homodyne detectors are sent to a data acquisition (DAQ) system for recording and post-processing. The quadrature correlations at specific sideband frequencies for different set phases $(\phi_i+\delta\theta_i$, $\theta_s)$ are used to demonstrate the frequency-dependent conditional squeezing. The phases $\phi_i$ and $\theta_s$ are actively stabilized to fix the quadrature detection in relation to the pump phase $\theta_{p}$.}
\label{fig:setup_methods}
\end{figure*}

As depicted in Extended Data Fig.~\ref{fig:setup_methods}, the vapour cell is illuminated by two optical fields. A circularly polarized optical repumping field, propagating transversely, prepares the spin ensemble in the hyperfine $|F=4, m_{F} = 4\rangle$ or $|F=4, m_{F} = -4\rangle$ ground state manifold with 82\% efficiency, enabling it to function as a macroscopic spin oscillator with an adjustable sign of the effective mass. 

The probe beam is blue-detuned by 1.6\,GHz from the D2 line $F =4 \rightarrow F' = 5$ transition to eliminate the absorption.  The polarization of the probe entering the atomic ensemble is chosen to optimize the light-spin interaction as discussed in the main text \cite{Thomas2021,Jia2023}.
The light-atoms interaction strength is characterized by the readout rate $\Gamma_{a} \propto g_{cs}^2 \mathcal{S}_{1}J_{x} \propto d$, where $g_{cs}$ is the single photon-atom coupling rate which depends on the probe detuning, and $\mathcal{S}_{1}$ is a Stokes parameter proportional to the power of the probe light. The optical depth of the spin ensemble, $d \propto J_{x} \propto N_{\text{Cs}}$, where $J_{x}$ is the macroscopic component of the collective spin, and $N_{\text{Cs}}$ is the number of atoms \cite{Jia2023}.

\begin{figure*}[htp]
\centering
\includegraphics[width = 0.7\textwidth]{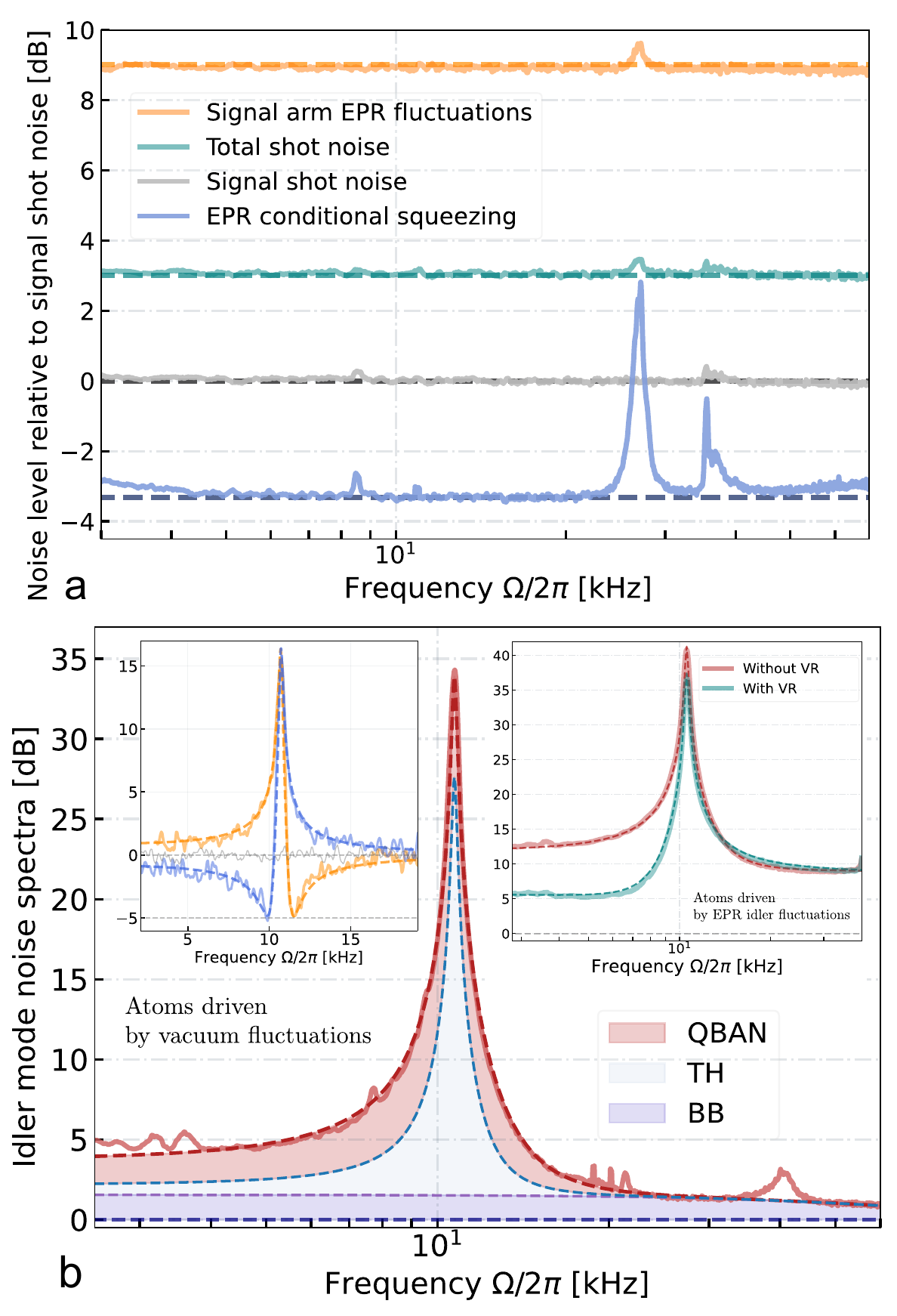}
\caption{\textbf{Calibration of the hybrid quantum system.} \textbf{a. Quantum noise of light fields in the absence of the atomic spin oscillator.} Teal trace -- total shot noise of the signal and idler. Gray trace -- shot noise of the signal. Blue trace shows $3.2$\,dB of conditional, frequency-independent EPR squeezing. It corresponds to $\simeq6$\,dB of entanglement of the EPR state when normalized to the total shot noise of the signal and idler fields (teal trace). Orange trace -- signal-arm EPR fluctuations. \textbf{b. Idler arm noise spectra in the presence of the atomic spin oscillator.} Main panel: Atoms driven by vacuum fluctuations; the reconstructed quantum-backaction noise (QBAN, red area), the decoherence-associated spin thermal (TH, light-blue area) and broadband noise (BB, purple area) are shown. Left inset: Ponderomotive squeezing of light by atoms observed for two different idler detection angles $\delta\theta_i$. Right inset: Atoms driven by EPR fluctuations; with (teal) and without (red) the virtual rigidity effect. Dashed curves are theoretical fits (see Methods for details). All traces are normalized to the idler shot noise. }
\label{fig:Hybrid system calibration}
\end{figure*}

The spin oscillator experiences depolarization, primarily due to the spontaneous emission that occurs in the presence of the probe field. This leads to the spin thermal noise imprinted onto the output probe beam. In the regime of strong light-atom coupling  $\Gamma_{a}\gg\gamma_{a}$, the quantum backaction induced by the probe dominates over the atomic thermal noise. 
Atoms that maintain interaction with the probe by multiple passages across the probe beam during the coherent evolution time contribute to the narrow-band atomic response limited by the spin decoherence rate $\gamma_{a}$. The remaining atoms contribute to faster-decaying atomic modes  leading to additional, broadband atomic noise with the bandwidth of $\gamma_{bb}\gg\gamma_{a}$ \cite{borregaard2016scalable,Shaham2020, Thomas2021}.  To minimise the broadband noise, the input field, comprising two orthogonally polarized fields (idler and $\text{LO}_{\text{i}}$), is shaped into a collimated square top-hat beam with a 7th-order super-Gaussian waist of 1.7\,mm by the top-hat shaper (THS) as shown in the Extended Data Fig.~\ref{fig:setup_methods}. It propagates through the cell with the filling factor $\sim 80 \%$ without introducing extra losses. The effect of the remaining broadband noise is illustrated in Extended Data Fig.~\ref{fig:Hybrid system calibration}. 

To characterise the spin oscillator, we block the idler EPR beam and drive the atoms with a coherent state of light. The Larmor frequency is set to $\Omega_{a}/2\pi \sim 10.5$\,kHz  and the quadrature of the output probe light is adjusted by a quarter- and a half-wave plate.  In Extended Data Fig.~\ref{fig:Hybrid system calibration}b, the recorded quantum noise, dominated by the spin quantum-backaction noise (QBAN) is shown as the red area. Choosing an optimal polarization of detected light we observe $5$\,dB of ponderomotive squeezing, as shown in the upper left inset of the figure. By fitting the displayed spin noise spectra, we can extract both the narrowband and broadband spin readout and decay rates along with the losses after the spin ensemble and the effective thermal occupancies (see Extended Data Table~\ref{tab:hybrid quantum network}). The total extracted occupancy number of $\approx 3.5$ is due to two factors: the imperfect spin polarization (measured using MORS) contributes $\approx 1$, and the technical noise sources contribute $\approx 2.5$ (for details see Supplementary Material Sec.~II~B). The effective thermal occupancies for the narrow band and broadband spin responses are the same. The reconstructed atomic thermal noise (the light blue area) and broadband noise (the purple area) are also presented in the Fig.~\ref{fig:Hybrid system calibration}b. The impact of these distinct spectral features on the conditional frequency-dependent squeezing level is discussed below. The red area, representing the quantum-backaction noise, highlights the key quantum contribution that enables the frequency-dependent rotation of the squeezing phase. The same calibration procedure was applied to experimental data at 54\,kHz Larmor frequency, demonstrated in Extended Data Fig.~\ref{fig:Hybrid system calibration}. The reduced classical noise at this frequency band allowed for better fitting. 

After characterizing the spin oscillator, we unblock the  idler output of the NOPO and record the dynamics of the spin oscillator driven by the idler component of the EPR field. The virtual rigidity phase $\delta \theta_{i}$ is extracted by fitting the noise spectrum shown in the right inset of Fig.~\ref{fig:Hybrid system calibration}b (green trace). By combining the overall losses obtained from the EPR source calibration with the optical and detection losses after atoms evaluated from ponderomotive squeezing, we estimate the propagation efficiency between the NOPO and atoms to be of $\eta_{i,\mathrm{in}} \sim 89 \%$. 
The optical losses from the atoms to the detection in the idler arm have been measured and result in the efficiency $\eta_{i,\mathrm{out}} \sim 90 \%$. Propagation efficiency from the NOPO and the detection efficiency in the signal arm result in $\eta_{s} \sim 92\%$. Based on those numbers, the predicted degree of two-mode squeezing fits the observed degree of squeezing as shown in Extended Data Fig.~\ref{fig:Hybrid system calibration}. 

Using the parameters extracted from these independent calibrations and the model presented in the next sections, Eq.~\eqref{mt:eq:OptimalSpectrum}, we calculate the predicted frequency-dependent conditional squeezing as a function of the signal homodyne angle $\theta_{s}$ shown in Figs.~\ref{fig:EntAtomsHybrid50kHzPosMass} and \ref{fig:EntAtomsHybrid10kHzPosNegMasVR} of the main text. The figures show good agreement between the predicted spectra of quantum noise and the experimental data.

\subsection*{Phase control of the signal and idler fields}

A feature of the EPR source critical for the present work is a phase control of the two-colour EPR state. Here we describe its underlying principles, while a complete report is in preparation \cite{UpComingEPR2}. The phases of the signal and idler beams of the NOPO are related to the pump phase by $\theta_p = \phi_i + \phi_s$ \cite{Reid1989,Schori2002}. To be able to precisely track the phase of the signal and the idler, we inject a coherent beam $\alpha_{c}$ co-propagating with the pump and frequency-shifted by $\delta\omega_c$ from the 1064\,nm laser by two acousto-optic modulators (AOMs). 

In this experiment we choose $\delta\omega_c /2\pi=3$\,MHz, well inside the NOPO cavity bandwidth, but also far from the resonance of the atomic spin oscillator.  An electronic reference phase for the control beam $\phi_c$ is provided by driving the AOMs by two outputs of a ultra-low phase noise direct digital synthesizer (DDS).

The control beam experiences the parametric interaction and is amplified while maintaining its phase, resulting in the output field $\alpha^{s}_{c}$. The interaction of control and pump provides the simultaneous generation of another coherent field, $\alpha^{i}_{c}$, centred at frequency $\omega_i-\delta\omega_c$. In this case, the field phase is the combination of the pump and signal control beam phases, $\phi_c^i=\theta_p-\phi_c$. The classical beams generated by this process propagate together with their respective entangled fields to the homodyne detections. We demodulate the photocurrents using the electronic reference. This provides an error signal proportional to the phase difference between the entangled fields and the local oscillators, measured by the two homodyne detectors (HDs). 
The photocurrent from HDs $I_s(t)$ contains the information from the signal-arm quadrature $q_s(\theta_s)$ and the beat note of the local oscillator with the control field shifted from the relevant signal band by $\delta\omega_c$ 
\begin{equation}
    I_s(t) \propto q_s(\theta_s) + \alpha^s_{c}\cos(\delta\omega_c t - \phi_c),
\end{equation}
with the second term allowing us to select and set the homodyne LO phase $\theta_{s}$. In turn, the outcome of the measurement of the idler field by means of the homodyne detection can be presented as 
\begin{equation}\label{mt:eq:I_i}
I_i(t) \propto Q_i(\theta_i) + \alpha^i_{c}\cos(\delta\omega_c t - \phi^i_c-\phi_i),
\end{equation}
where the first term is the idler quantum field contribution and second term is the beat note between the second control field which has passed through the NOPO channel and the respective local oscillator.  The demodulated signal at $\delta\omega_c/2\pi$ provides a tool to control the homodyne phase $\theta_{i}=\phi_{i}+\delta\theta_{i}$, whereas the phase offset $\delta\theta_{i}$ can be tuned separately using the quarter-wave plate (see the setup in Extended Data Fig.~\ref{fig:setup_methods} and Fig.~\ref{fig:1_motivation_experiment} in the main text). An exact definition of $Q_i(\theta_i)$ is given in the next section.

The frequency offset of $\delta \omega_c/2\pi\simeq 3$\,MHz, guarantees that locking of those phases is not affected by the atomic spin oscillator. The procedure described above provides a set of well-defined phases that we use as references for the scan of $\theta_s$ involved in the demonstration of frequency-dependent conditional squeezing.

\subsection*{Idler field and light-atoms interaction}

The 1.6\,GHz detuning of the 852\,nm NOPO idler field from atomic resonance is achieved by changing the frequency of the 1064\,nm laser, which alters the NOPO cavity length and its resonance condition. The 852\,nm laser lock follows the change, enabling fine-tuning of the idler field with precision close to the NOPO cavity bandwidth. 

To enable interaction between the idler field and the atomic spin ensemble, we overlap the idler output of the NOPO with an orthogonally polarized probe beam. This is realised by combining the linearly polarised idler and probe beams on a Polarising Beam Splitter (PBS), which transforms the idler field’s quadrature fluctuations into Stokes operator fluctuations \cite{Luis2006,Moller2017,bowen2002polarization}.
For a linearly polarised, strong coherent probe beam with a relative phase $\phi_i$ between the idler and the probe, the quadratures map onto the Stokes operators as ${q}_i(\phi_i) = \sqrt{2}\mathcal{S}_{2}^{\mathrm{in}}/|\alpha_\mathrm{pr}|$ and ${q}_i(\phi_i + \pi/2) = \sqrt{2}\mathcal{S}_{3}^{\mathrm{in}}/|\alpha_\mathrm{pr}|$, where $\alpha_\mathrm{pr}$  is the amplitude of the probe field, $\{ \mathcal{S}_{0},\mathcal{S}_{1},\mathcal{S}_{2},\mathcal{S}_{3} \}$ are quantum Stokes operators, and ${q}_{i}(\phi_{i}) = \cos(\phi_{i}) {p}_{i} + \sin(\phi_{i}) {x}_{i}$ is the quadrature of the optical field entering the atomic ensemble.

The idler field encoded into the polarisation state is processed by the atomic ensemble. The output Stokes parameters are then given by $ \mathcal{S}_{3}^{\text{out}} = \mathcal{S}_{3}^{\text{in}}$ and $\mathcal{S}_{2}^{\text{out}} = \mathcal{S}_{2}^{\text{in}} + \mathcal{K}_{a} \mathcal{S}_{3}^{\text{in}} +\mathcal{N}_a$, where $\mathcal{N}_a=N_a|\alpha_\mathrm{pr}|/\sqrt{2}$; $\mathcal{K}_{a}$ and $N_a$ are defined in the main text.

The light emerging from the spin ensemble is measured by the polarization homodyne detection. The diagonal linear and circular polarization operators $\mathcal{S}_{2}$ and $\mathcal{S}_{3}$ are measured by passing light through a half-wave plate, and an additional quarter-wave plate (QWP), respectively, followed by the PBS, as shown in the Extended Data Fig.~\ref{fig:setup_methods}.
The resulting measured Stokes quadrature operator is  $\mathcal{R}_{i}(\delta\theta_i)  = \mathcal{S}_{2}^{\text{out}} \cos(\delta \theta_i) + \mathcal{S}_{3}^{\text{out}} \sin(\delta \theta_i)= Q_{i}(\delta\theta_i) |\alpha_\mathrm{pr}|/\sqrt{2}$, where the phase $\delta \theta_{i}$ is set by the orientation of the QWP. 

The phase $\phi_i$ is set by a coherent control loop that monitors the phase offset between the probe beam and the control field at the idler homodyne detector. The control field follows the same propagation path as the idler beam, but is unaffected by the atomic oscillator due to a frequency offset, $\delta\omega_\mathrm{c}$. The phase offset $\phi_{i}$ measured by the coherent control loop combines the encoding phase and $\delta\theta_{i}$.

Throughout those measurements, we set $\theta_{i} = 0$, so that the  quadrature of the idler detected outside the bandwidth of the spin dynamics bandwidth ($\Omega \gg |\Omega_a|$) is ${p}_{i}$. In the setting not engaging virtual rigidity, that is in the absence of QWP ($\delta\theta_{i} = 0$), the coherent control loop sets the encoding phase $\phi_{i} = 0$, corresponding to $\mathcal{S}_{2}^{\text{in}} = {x}_{i} |\alpha_\mathrm{pr}|/\sqrt{2}$ and $\mathcal{S}_{3}^{\text{in}} = {p}_{i}|\alpha_\mathrm{pr}|/\sqrt{2}$. With such settings, the final idler photocurrent within  the relevant range of frequencies $\Omega$ is proportional to ${P}_{i} = {p}_{i} + \mathcal{K}_{a}(\Omega){x}_{i} + N_a(\Omega)$. For general $\delta\theta_i$, we have $Q_i(\delta\theta_i) = X_i\sin(\delta\theta_i) + P_i\cos(\delta\theta_i)$.

\subsection*{Optimal quantum noise reduction by Wiener filtering in the idler channel}

The squeezing we demonstrate for the signal arm is conditioned on the idler-arm measurement
\begin{equation}
Q_{s|i}(\Omega,\theta_s)=q_{s}(\Omega,\theta_s)+g(\Omega,\theta_s,\delta\theta_{i})Q_i(\Omega,\delta\theta_{i}).
\end{equation}
To achieve the maximal frequency-dependent squeezing of the signal field based on the idler measurement record, a Wiener filter $g(\Omega,\theta_s,\delta\theta_{i})$ is designed for each detection phase $\theta_s$, which provides an optimal estimate of the correlated quantum noise in the idler channel. By subtracting the filtered idler quantum noise from the signal noise, the optimal reduction of the signal quantum noise limited by the remaining uncorrelated noise is achieved \cite{Danilishin2019, Ma2017,gould2021optimal}.

Our protocol is compatible with both causal and non-causal filtering $g(\Omega,\theta_s,\delta\theta_{i})$ of the idler measurement record. Causal filtering, which uses only past idler measurements to reduce the signal arm noise at a given time, is required for using the sensor for, e.g., real-time signal tracking or adaptive sensing. On the other hand, the non-causal filtering uses the full idler measurement record and is relevant for sensing scenarios where the signal is extracted by post-processing of the full measurement record.

In this work we focus on non-causal conditioning, thereby emphasizing the maximally achievable squeezing.
In this case, the optimal gain is given by the idler spectral density $S_{Q_{i}}$ and the cross-spectral density $S_{q_{s},Q_{i}}$ of the two entangled channels as 
\begin{equation}
    g(\Omega,\theta_s,\delta\theta_{i}) = - \frac{S_{q_{s},Q_{i}}(\Omega,\theta_s,\delta\theta_{i})}{S_{Q_{i}}(\Omega,\delta\theta_{i})}\label{mt-eq:WienerGain}
\end{equation}
which leads to the power spectral density of the optimised signal 
\begin{equation}
    S_{Q_{s|i}}(\Omega,\theta_s) = S_{q_{s}}(\Omega,\theta_s) - \frac{|S_{q_{s},Q_{i}}(\Omega,\theta_s,\delta\theta_{i})|^2}{S_{Q_{i}}(\Omega,\delta\theta_{i})}. \label{mt-eq:Opt_signal}
\end{equation}

The optimal filter~\eqref{mt-eq:WienerGain} automatically takes into account the signal homodyne detection phase $\theta_{s}$, deleterious noise introduced in the idler path, and phase shifts due to the atomic dissipation through decoherence, as represented by the complexity of the response function $\mathcal{K}_a(\Omega)$ [see Eq.~\eqref{eq:K_a} in the main text]. Additionally, the Wiener filter can potentially compensate for imperfect matching between the idler and the quantum sensor response introduced in the signal arm in a particular sensing application. 
The analysis of the experimental data using this approach is detailed in the Supplementary Material Sec.~II.

\subsection*{Theory of frequency-dependent conditional squeezing}

Figs.~\ref{fig:EntAtomsHybrid50kHzPosMass} and \ref{fig:EntAtomsHybrid10kHzPosNegMasVR} in the main text present comparisons of the measured squeezing with the theoretical model, which we describe here. The model consists of equations of motion of the spin oscillator and input-output relations describing its interaction with light. It accounts for contributions of thermal and broadband noise to the response of the spin system, as well as for the effects of imperfect readout efficiency ($\eta_{i,\text{out}} < 1$), imperfect coupling of the idler beam to the oscillator ($\eta_{i,\text{in}} < 1$) and the readout efficiency in the signal arm, $\eta_{s} < 1$. From this model we obtain an expression for the power spectral density of the conditioned signal photocurrent ${Q}_{s|i}$, which, in turn, is minimized using Wiener filter theory as outlined in the preceding Methods subsection. Detailed calculations are provided in the Supplementary Material Sec.~I~C. 

Here we present a general formula for an arbitrary phase offset $\delta\theta_{i}$ imposed by the quarter-wave plate and giving rise to virtual rigidity. The case of no virtual rigidity can be obtained by setting $\delta\theta_{i}=0$. 
The spectrum of the optimally conditioned signal, normalised to the signal shot noise $S_{SN}$ level, in the signal-arm detection quadrature $\theta_s$ is given by
\begin{equation}
\begin{aligned}
    \frac{S_{{Q}_{s|i}}}{S_{SN}} &= 1-\eta_{s} + \frac{\eta_{s}}{\cosh(2r)}\times \\
&\Bigg[
\cosh^2(2r) -
   \frac{
    \sinh^2(2r)
   \left|\cos(\theta_{s}) - \sin(\theta_{s}) \mathcal{K}^{\text{eff}}_{\text{a}} \right|^2
    }{
    1+ \left|\mathcal{K}^{\mathrm{eff}}_{a}\right|^2
    +2\tfrac{\Lambda^{\text{eff}}_{\text{in}} + \Lambda^{\text{eff}}_{\text{out}} + 
        {S}^{\text{eff}}_{\text{th}}
         +{S}^{\text{eff}}_{\text{bb}}
    }{\eta_{i,\text{in}} \cosh(2r)}
    }
\Bigg], \label{mt:eq:OptimalSpectrum}
\end{aligned}
\end{equation}
where dependencies on $\Omega$ are omitted for brevity. The numerator of the second term in the brackets represents the correlation between the signal and idler, shaped by the backaction of the atomic spin oscillator. The denominator captures the spectrum of the idler signal, incorporating, in addition to the backaction, the following four terms, responsible for the various deleterious effects mentioned above,
\begin{align}
    \Lambda^{\text{eff}}_{\text{in}}(\Omega)     &= \frac{1-\eta_{i,\text{in}}}{2}(1+ |\mathcal{K}^{\text{eff}}_{\text{a}}(\Omega)|^2) \\
    \Lambda^{\text{eff}}_{\text{out}}(\Omega)    &= \frac{1-\eta_{i,\text{out}}}{2\eta_{i,\text{out}}} \frac{1}{|g_{\mathrm{VR}}(\Omega)|^2}  \\ 
    {S}^{\text{eff}}_{\text{th}}(\Omega) &= \left|\mathcal{K}_\mathrm{th}(\Omega)\frac{\cos(\delta\theta_{i})}{g_{\mathrm{VR}}(\Omega)}\right|^2  (1/2+n_{\text{th}}) \\ 
    {S}^{\text{eff}}_{\text{bb}}(\Omega) &= \left|\mathcal{K}_\mathrm{bb}(\Omega)\frac{\cos(\delta\theta_{i})}{g_{\mathrm{VR}}(\Omega)}\right|^2  (1/2+n_{\text{bb}}).
\end{align}
In order of appearance, they represent the effects of 
suboptimal coupling of the idler field to the spin ensemble, $\eta_{i,\text{in}}$, 
imperfect readout efficiency of the idler detector, $\eta_{i,\text{out}}$, thermal occupation of the collective spin state, $n_{\text{th}}$, governed by $\mathcal{K}_{th}(\Omega) = \sqrt{2\gamma_{a}\Gamma_{a}}\Omega_{a}/(\Omega^{2}_{a} - \Omega^{2} - i \gamma_{a} \Omega + \gamma_{a}^{2} / 4)$ and the broadband noise occupation number, $n_{\text{bb}}$, governed by $\mathcal{K}_{bb}(\Omega) = \sqrt{2\gamma_{bb}\Gamma_{bb}}\Omega_{a}/(\Omega^{2}_{a} - \Omega^{2} - i \gamma_{bb} \Omega + \gamma_{bb}^{2} / 4)$. The effect of the \emph{virtual rigidity} is captured by the gain
\begin{equation}
        g_{\mathrm{VR}}(\Omega) = 1-\mathcal{K}_{a}(\Omega, \Omega_{a},\Gamma_{a})\frac{\sin(2\delta\theta_{i})}{2}.    
\end{equation}
The effective backaction $\mathcal{K}_{a}^{\text{eff}}(\Omega)\equiv \mathcal{K}_{a}(\Omega, \Omega_{a}^{\text{eff}},\Gamma_{a}^{\text{eff}})$ is defined as the backaction coefficient used in Eq.~\eqref{eq:K_a} of the main text, but evaluated with the effective readout rate and the effective Larmor frequency
\begin{align}
        \Gamma_{a}^{\text{eff}} &= \Gamma_{a}\frac{\cos^{2}(\delta\theta{i})}{\sqrt{1-\frac{\Gamma_{a}}{2\Omega_{a}}\sin(2\delta\theta_{i})}} \label{eq:GammaEff} \\
        \Omega_{a}^{\text{eff}} &= \Omega_{a}\sqrt{1-\frac{\Gamma_{a}}{2\Omega_{a}}\sin(2\delta\theta_{i})};  \label{eq:LarmEff}
\end{align}
these expressions are valid insofar as the quantity appearing under the square roots is positive, as is the case for our system parameters.

By minimizing Eq.~\eqref{mt:eq:OptimalSpectrum} as a function of $\theta_s$ for each Fourier component $\Omega$ separately, we find the optimal angle $\theta_s=\Phi_\mathrm{VR}(\Omega)$ to be the solution to the set of equations
\begin{subequations}\label{mt-eq:Phi_VR-general}
\begin{align}
    \tan(2\Phi_\mathrm{VR}(\Omega))
    &=\frac{-2\text{Re}[(\mathcal{K}_a^{\mathrm{eff}}(\Omega))^{-1}]}{|(\mathcal{K}_a^{\mathrm{eff}}(\Omega))^{-1}|^{2}-1}\\
    \text{sign}[\cos(2\Phi_\mathrm{VR}(\Omega))]&=\text{sign}(1-|\mathcal{K}^{\mathrm{eff}}_{a}(\Omega)|^{2})
\end{align}
\end{subequations}
In the limit $\gamma_{a}^2\ll(\Omega_{a}^{\text{eff}})^2,(\Gamma_{a}^{\text{eff}})^2$ we have $|(\mathcal{K}_a^{\mathrm{eff}}(\Omega))^{-1}|^2 \approx \text{Re}^2[(\mathcal{K}_a^{\mathrm{eff}}(\Omega))^{-1}]\approx [(\mathcal{K}_a^{\mathrm{eff}}(\Omega))^{-1}]^2_{\gamma_a\rightarrow 0}$, whereby Eqs.~\eqref{mt-eq:Phi_VR-general} reduce to 
\begin{equation}\label{mt-eq:Phi_VR-approx}
\Phi_\mathrm{VR}(\Omega) \approx - \arctan(\mathcal{K}_{a}^{\mathrm{eff}}(\Omega))|_{\gamma_a\rightarrow 0}
\end{equation}
as presented in the main text. 
This expression, valid for general $\delta\theta_i$, is used to generate the dashed curves presented as $\Phi_j(\Omega)$ in the main text Figs.~\ref{fig:EntAtomsHybrid50kHzPosMass}b and \ref{fig:EntAtomsHybrid10kHzPosNegMasVR}(b,d,f) for the various special cases labelled $j\in\{\pm,\mathrm{VR}\}$. Evaluating Eq.~\eqref{mt:eq:OptimalSpectrum} at $\theta_s=\Phi_\mathrm{VR}(\Omega)$ yields the achieved degree of (frequency-dependent) squeezing; this amounts to replacing the factor 
\begin{multline}
|\cos(\theta_{s}) - \sin(\theta_{s}) \mathcal{K}^{\text{eff}}_{\text{a}} |^2\rightarrow \\
\frac{1+|\mathcal{K}_{a}^{\mathrm{eff}}|^{2}}{2}\left[1+\sqrt{1-\frac{4\text{Im}^{2}[\mathcal{K}_{a}^{\mathrm{eff}}]}{\left(1+|\mathcal{K}_{a}^{\mathrm{eff}}|^{2}\right)^{2}}}\right]\\
\approx1+ |\mathcal{K}^{\text{eff}}_{\text{a}}|^2, 
\end{multline}
where the approximation is valid in the limit $\gamma_{a}^2\ll(\Omega_{a}^{\text{eff}})^2,(\Gamma_{a}^{\text{eff}})^2$. 

In the aforementioned limit, we may use Eqs.~\eqref{eq:K_a} and \eqref{mt-eq:Phi_VR-approx} to derive an expression for the bandwidth $\delta\Omega_\mathrm{SQL}>0$ over which the rotation angle $\Phi(\Omega)$ ($\in[0,\pi]$ for specificity) changes by $45^\circ$ relative to its value at the effective spin oscillator resonance $\Phi(\Omega_a^\mathrm{eff})=\pi/2$, i.e., that obeys $|\Phi(\Omega_a^\mathrm{eff}+\delta\Omega_\mathrm{SQL})-\Phi(\Omega_a^\mathrm{eff})|=\pi/4$. The result is $\delta\Omega_\mathrm{SQL}=|\Omega_{a}^{\mathrm{eff}}|(\sqrt{1+\Gamma_{a}^{\mathrm{eff}}/|\Omega_{a}^{\mathrm{eff}}|}-1)\approx\Gamma_{a}^{\mathrm{eff}}/2$, where the approximation holds under the additional assumption $\Gamma_{a}^{\mathrm{eff}}\ll |\Omega_{a}^{\mathrm{eff}}|$.

In addition to comparing our model to the squeezing achieved in the present experiment, we also used the model to predict the degree of broadband noise reduction obtained by reducing the two main imperfections of our system: broadband spin noise and thermal spin noise. The results are reported in Extended Data Fig.~\ref{fig:conditional squeezing with improved system}, while further details can be found in the Supplementary Material Sec.~II~B. 


\section*{Acknowledgments}
This work has been supported by VILLUM FONDEN under a Villum Investigator Grant, grant no.\ 25880, by the Novo Nordisk Foundation through Copenhagen Center for Biomedical Quantum Sensing, grant number NNF24SA0088433 and through ‘Quantum for Life’ Center, grant NNF20OC0059939, by the European Union's Horizon 2020 research and innovation program under the Marie Sklodowska-Curie grant agreements No.\ 125101 `EPROXY' and No.\ 122436 QNOIWA, and the European Research Council Advanced grant QUANTUM-N. We acknowledge Alkiviadis Zoumis and Ryan Yde for their experimental contributions. We acknowledge Farid Khalili for insightful discussions and careful reading of the manuscript. We acknowledge valuable discussions with Jürgen Appel, Mikael Lassen, Rodrigo Thomas and Micha\l{} Parniak at the early stages of this project.  


\section*{Author Contributions}
J.J.\ and Mm.B.\ developed the atomic spin oscillator setup.
T.B.B.\ designed the two-colour EPR source and built it together with V.N.\ and A.G..
V.N., J.J., T.B.B., A.G.\ and Mm.B.\ performed the data acquisition.
E.Z., V.N., J.J.\ and A.G.\ developed the analytical model.
J.J.\ and V.N.\ performed the data analysis.
Mk.B.\ fabricated the caesium cell.
J.H.M.\ contributed with valuable discussions.
T.B.B.\ led the experimental work. 
V.N., J.J., T.B.B. and A.G.\ contributed equally to this work.
E.S.P.\ conceived and led the project.
All authors contributed to writing the paper.


\section*{Author Information}
The authors declare no competing financial interests. Correspondence and requests for materials should be addressed to polzik@nbi.ku.dk.


\section*{Data Availability Statement}
The data supporting this study's findings are available at ERDA.

\newpage

\begin{figure*}[htp]
\centering
\includegraphics[width=0.7\textwidth]{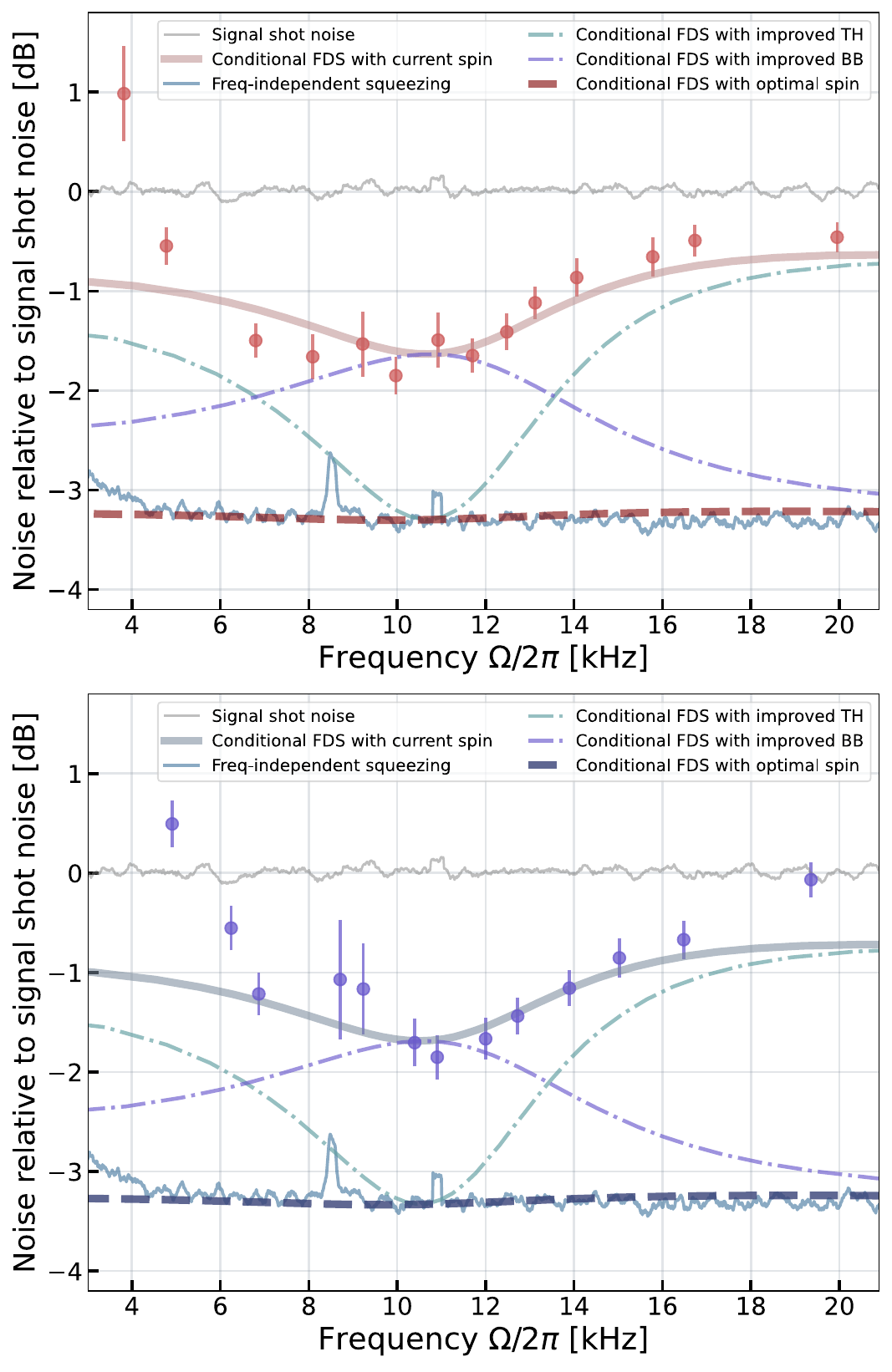}
\caption{\textbf{Demonstration of overall quantum noise reduction with potential improvements to the spin oscillator.} Red (positive mass) and blue (negative mass) points show the measured minimal conditional squeezing at each signal phase, along with the overall modelled quantum noise reduction based on the calibrated parameters of the current spin system. The discrepancy between theoretical predictions and experimental data at frequencies below 5\,kHz is attributed to residual laser amplitude noise. Estimated conditional squeezing level are also presented for two scenarios: when the atomic thermal (TH) occupation is reduced by a factor of 3 (purple curve) and when the atomic broadband (BB) readout rate is reduced by a factor of 6 (teal curve) relative to current conditions. With both noise sources reduced, the optimal spin system generates the conditional frequency-dependent squeezing (FDS) (red/blue dashed line), limited only by the frequency-independent level of squeezing of light (light blue trace). 
}
\label{fig:conditional squeezing with improved system}
\end{figure*}
\begin{figure*}[htp]
\centering
\includegraphics[width=0.9\textwidth]{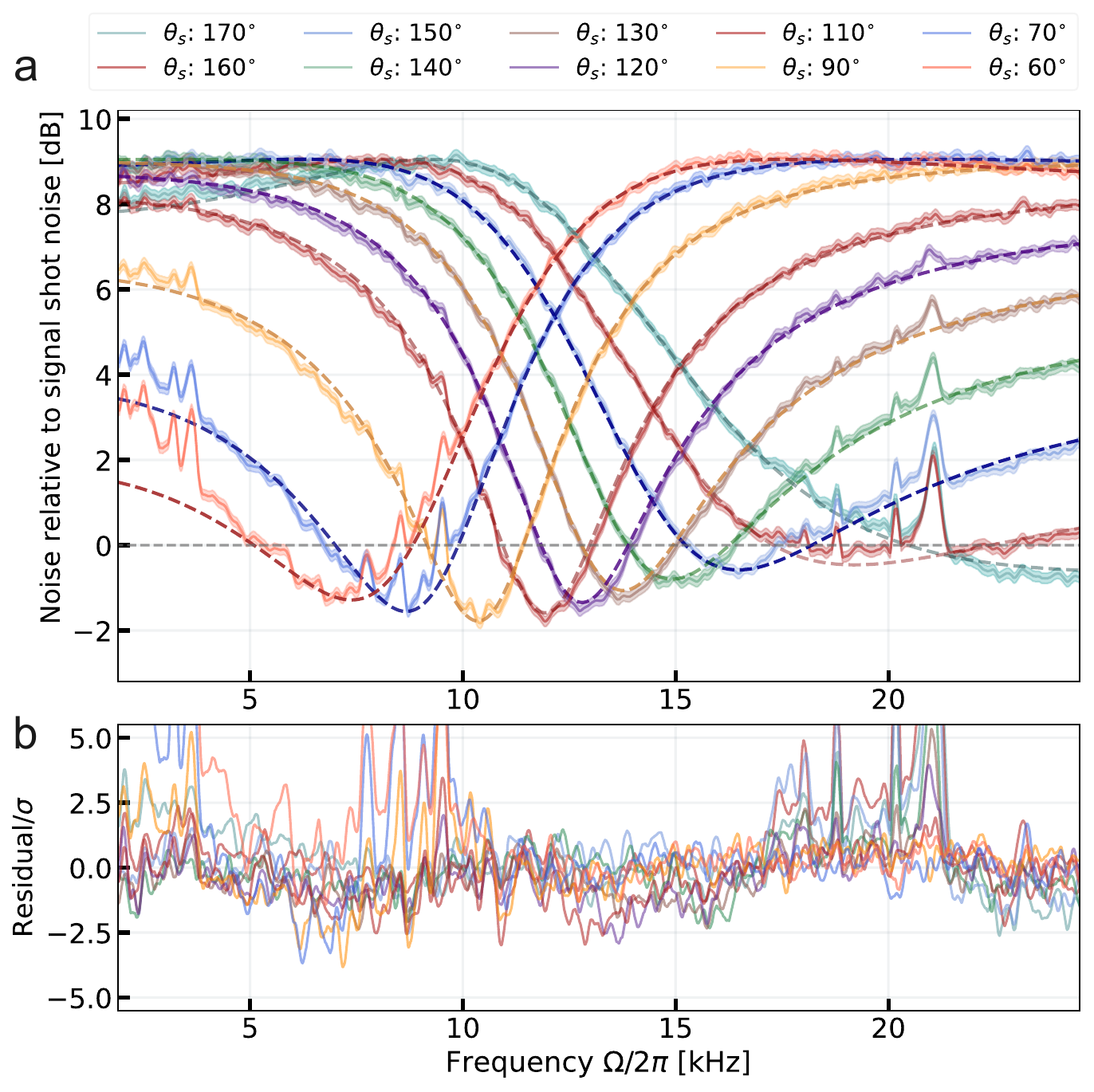}
\caption{\textbf{Uncertainty analysis for frequency-dependent conditional squeezing via an effective positive-mass oscillator $\Omega_a/2\pi=10.7\,\mathrm{kHz}$ at different signal homodyne detection phases (10 out of 18).} \textbf{a}. The solid lines represent the inferred quantum noise on the signal detector conditioned on the idler measurement, with the shaded regions indicating the 1$\sigma$ uncertainty ($68\%$ confidence interval). The dashed lines
show the quantum noise model fitted to the measured data using a $\chi^{2}$ method. \textbf{b}. The residuals between the measured data and the fitted model, normalized to 1$\sigma$ uncertainty is plotted. Noise spikes near 20\,kHz (attributed to the etalon dither lock of the 852\,nm laser) and classical noise (from both laser intensity noise and magnetic noise sources) at low acoustic frequency (below 10\,kHz) are clearly visible in the residual plots.} 
\label{fig:conditional squeezing with positive mass}
\end{figure*}

\begin{figure*}[htp]
\centering
\includegraphics[width=0.9\textwidth]{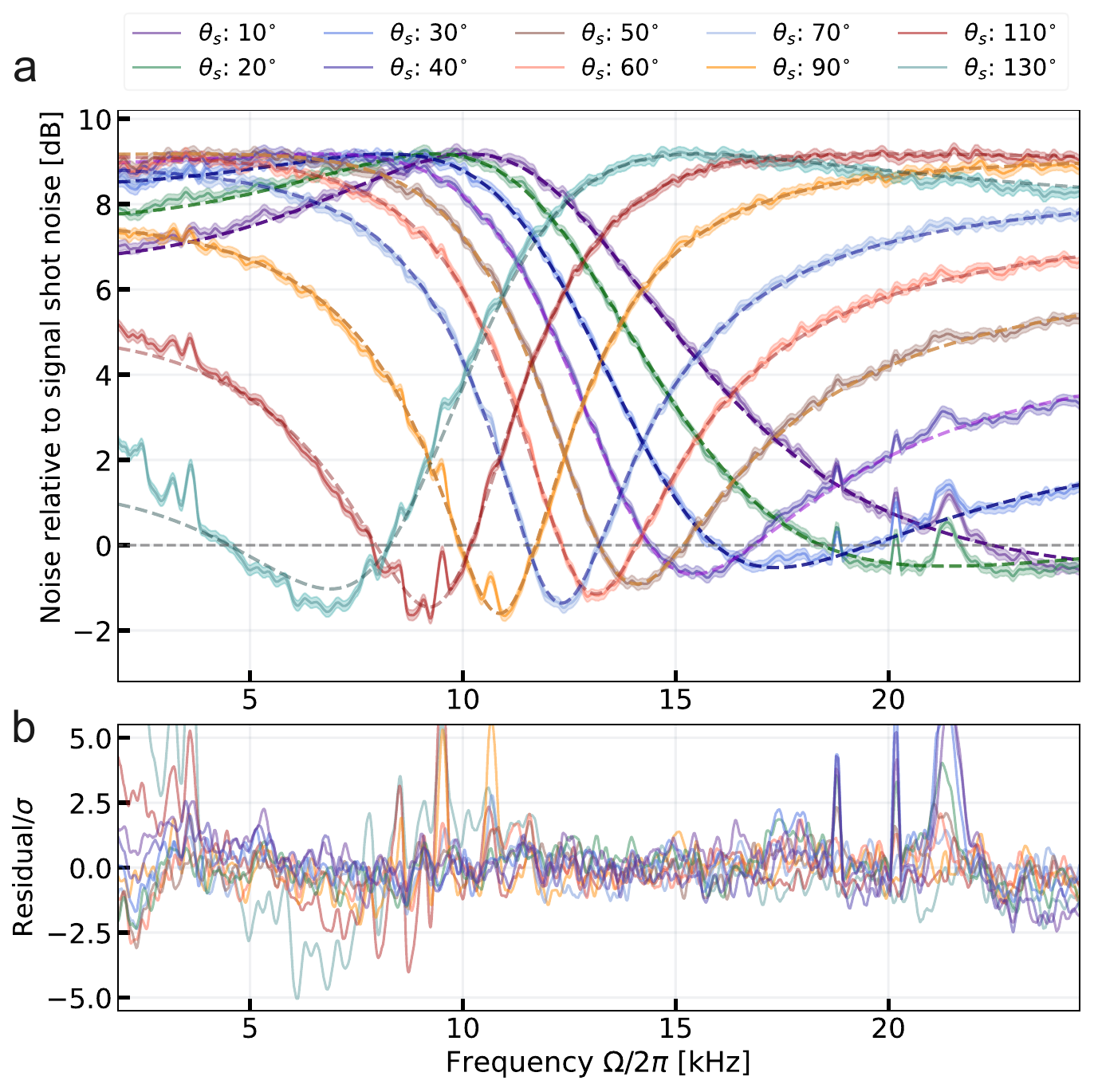}
\caption{\textbf{Uncertainty analysis for frequency-dependent conditional squeezing via an effective negative-mass oscillator $\Omega_a/2\pi=-10.5\,\mathrm{kHz}$ at different signal homodyne detection phases (10 out of 18).} \textbf{a}. The solid lines represent the inferred quantum noise on the signal detector conditioned on the idler measurement, with the shaded regions indicating the 1$\sigma$ uncertainty ($68\%$ confidence interval). The dashed lines
show the quantum noise model fitted to the measured data using a $\chi^{2}$ method. \textbf{b}.  The residuals between the measured data and model fit, normalized to 1$\sigma$ uncertainty are also presented.
}
\label{fig:conditional squeezing with negative mass}
\end{figure*}

\begin{table*}[h!]
    \centering
    \renewcommand\cellalign{lc} 
    \begin{tabular}{>{\raggedright\arraybackslash}p{8.5cm} >{\raggedright\arraybackslash}p{3cm} >{\raggedleft\arraybackslash}p{3cm}}
        \toprule
        \textbf{Parameter} & \textbf{Symbol} & \textbf{Value} \\ 
        \midrule
        \multicolumn{3}{c}{\textbf{Hybrid system \& detection}} \\ 
        \midrule
        Two-mode squeezing factor & $r$ & 1.42 | 1.42 | 1.31 \\
        Propagation efficiency before atoms  & $\eta_{i,\mathrm{in}}$ & 0.89 \\
        Overall efficiency after atoms  & $\eta_{i,\mathrm{out}}$ & 0.90 \\
        Signal-arm net efficiency (propagation \& detection) & $\eta_{s}$ & 0.92 \\
        Idler-arm net efficiency (propagation \& detection) & $\eta_{i}\,(=\eta_{i,\mathrm{out}}\eta_{i,\mathrm{in}})$ & 0.8 \\
        QWP phase  & $\delta\theta_{i}$ &  0$^{\circ}$,  42$^{\circ}$ \\
        Signal LO power &  & 1 mW \\
        Idler LO power &  & 1 mW \\
        \midrule
        \multicolumn{3}{c}{\textbf{Atomic spin oscillator}} \\
        \midrule
        Effective spin oscillator  mass &   & p | n | p\\
        Larmor frequency & $\Omega_{a}/2\pi$ & 10.7 | 10.5 | 54 kHz \\
        Spin readout rate & $\Gamma_{a}/2\pi$ & 9.3 | 9.5 | 8.5 kHz \\
        Spin decoherence rate & $\gamma_{a}/2\pi$ & 240 | 240 | 200 Hz \\
        Effective spin thermal occupation & $n_{th}$ &  3.5 | 3.4 | 4.0 \\
        Broadband noise in SN unit at $\Omega_{a}$  &  &  1.8 | 1.9 | 1.3  dB \\
        Spin Broadband readout rate & $\Gamma_{bb}/2\pi$ & 140 | 130 | 5  kHz\\
        Spin Broadband decoherence rate & $\gamma_{bb}/2\pi$ & 190 | 190 | 190 kHz\\
        Effective spin broadband occupation & $n_{bb}$ &  3.5 | 3.4 | 4.0 \\
        Probe field detuning & $\Delta_{a}/2\pi$ & 1.6 GHz\\
        Probe input polarization & $\alpha$ & 45$^{\circ}$\\
        Spin polarization &  & ~82\%\\
        \bottomrule
    \end{tabular}
    \caption{Summary of notations and experimental parameters for the hybrid quantum system.  p | n stands for the positive/negative-mass configuration. The values are estimated by $\chi^2$ fits of the calibration data.}
    \label{tab:hybrid quantum network}
\end{table*}

\end{document}


\title{Supplementary Material for\\``Hybrid quantum network for sensing in the acoustic frequency range''}

\author{Valeriy Novikov}
\thanks{These authors contributed equally}
\affiliation{Niels Bohr Institute, University of Copenhagen, Blegdamsvej 17, DK-2100 Copenhagen Ø, Denmark}
\affiliation{Russian Quantum Center, Skolkovo, Moscow, Russia}

\author{Jun Jia}
\thanks{These authors contributed equally}
\affiliation{Niels Bohr Institute, University of Copenhagen, Blegdamsvej 17, DK-2100 Copenhagen Ø, Denmark}

\author{Túlio Brito Brasil}
\thanks{These authors contributed equally}
\affiliation{Niels Bohr Institute, University of Copenhagen, Blegdamsvej 17, DK-2100 Copenhagen Ø, Denmark}

\author{Andrea Grimaldi}
\thanks{These authors contributed equally}
\affiliation{Niels Bohr Institute, University of Copenhagen, Blegdamsvej 17, DK-2100 Copenhagen Ø, Denmark}

\author{Maimouna Bocoum}
\affiliation{ESPCI Paris, PSL University, CNRS, Institut Langevin, Paris 75005, France}
\affiliation{Niels Bohr Institute, University of Copenhagen, Blegdamsvej 17, DK-2100 Copenhagen Ø, Denmark}

\author{Mikhail Balabas}
\affiliation{Niels Bohr Institute, University of Copenhagen, Blegdamsvej 17, DK-2100 Copenhagen Ø, Denmark}

\author{J\"{o}rg Helge M\"{u}ller}
\affiliation{Niels Bohr Institute, University of Copenhagen, Blegdamsvej 17, DK-2100 Copenhagen Ø, Denmark}

\author{Emil Zeuthen}
\affiliation{Niels Bohr Institute, University of Copenhagen, Blegdamsvej 17, DK-2100 Copenhagen Ø, Denmark}

\author{Eugene Simon Polzik}
\affiliation{Niels Bohr Institute, University of Copenhagen, Blegdamsvej 17, DK-2100 Copenhagen Ø, Denmark}

\begin{abstract}
In this supplementary material, we detail the theoretical foundations of our work. Specifically, we develop the theory of frequency-dependent conditional squeezing enabled by an atomic spin oscillator and an entangled light source using various quantum noise engineering approaches.
Additionally, we define the (optimal) Wiener filter for inferring conditional squeezing and outline our practical implementation of the filter.
Finally, based on the calibration of the hybrid quantum network performance, we assess the limitations of the current hybrid setup and propose strategies for future improvement.
\end{abstract}


\maketitle

\tableofcontents

\newpage

\section{Theoretical model}

In this section, we present the theoretical model used to analyse the experimental data and extract essential parameters for our system. The main results are reported in the Methods; here we present their detailed derivations.

\subsection{Spin oscillator in the virtual-rigidity configuration}\label{sc:VRmodel}

The so-called virtual-rigidity configuration is obtained by inserting a quarter-wave plate before the idler homodyne detector in order to generate a phase shift $\delta\theta_{i}\neq 0$, see Extended Data Fig.~\ref{fig:setup_methods}.  In this case, the encoding phase changes to $\phi_{i} = -\delta\theta_{i}$, and the Stokes operators become
\begin{subequations}\label{apx:eq:VRencoding}
\begin{align}
    \hat{\mathcal{S}}_{3}^{\text{in}}\sqrt{2}/\alpha_\textrm{pr} &= -\hat{p}_{\text{i}} \sin(\delta\theta_{i}) + \hat{x}_{\text{i}}\cos(\delta\theta_{i}),\\
    \hat{\mathcal{S}}_{2}^{\text{in}}\sqrt{2}/\alpha_\textrm{pr} &=  \hat{p}_{\text{i}} \cos(\delta\theta_{i}) + \hat{x}_{\text{i}}\sin(\delta\theta_{i}).
\end{align}
\end{subequations}
The backaction on the atomic spin oscillator, driven by $\hat{\mathcal{S}}_{3}^{\text{in}}$, is now defined by a combination of the phase and amplitude quadratures of the idler.
The second effect of the quarter-wave plate is the change of the detected quadrature, which becomes proportional to a linear combination of output Stokes operators, $\hat{\mathcal{S}}^{\text{out}}_{2}$ and $\hat{\mathcal{S}}_{3}^{\text{out}}$, 
\begin{equation}
    \hat{\mathcal{R}}_{i}(\delta\theta_{i}) = \left( \hat{\mathcal{S}}_{2}^{\text{out}}\cos(\delta\theta_{i}) + \hat{\mathcal{S}}_{3}^{\text{out}} \sin(\delta\theta_{i}) \right) = \frac{|\alpha_\textrm{pr}|}{\sqrt{2}} \hat{Q}_{i}(\delta\theta_{i})
    \label{apx:eq:IdlerStocksVR}.
\end{equation}
Applying  Eqs.~\eqref{apx:eq:VRencoding} to the detected quadrature Eq.~\eqref{apx:eq:IdlerStocksVR} and using Eq.~\eqref{eq:IO-P_i} of the main text, we can rewrite the idler photocurrent as a function of the idler canonical quadratures:
\begin{align}
   \hat{Q}_{i}(\delta\theta_{i})  &= \hat{p}_{i} 
    + \left( \mathcal{K}_{a}(\Omega) \hat{a}_{i}^{\pi/2-\delta\theta_{i}} +\hat{N}_a(\Omega) \right) \cos(\delta\theta_{i}),   
    \label{apx:eq:QuadraturewQW_wCC}
\end{align}
where $\hat{a}_{i}^{\pi/2-\delta\theta_{i}} = -\hat{p}_{\text{i}} \sin(\delta\theta_{i}) + \hat{x}_{\text{i}}\cos(\delta\theta_{i})$, is a generic input quadrature defined by the encoding phase $\phi_{i} = -\delta\theta_{i}$. According to Ref.~\cite{Zeuthen2019}, this equation describes an oscillator with an effective read-out rate $\Gamma_{a}^{\text{eff}}$ and an effective resonance frequency $\Omega_{a}^{\text{eff}}$; in terms of the quantities used in the present work, we obtain
\begin{align}
        \Gamma_{a}^{\text{eff}} &= \Gamma_{a}\frac{\cos^2(\delta\theta_{i})}{\sqrt{1-\tfrac{\Gamma_{a}}{2\Omega_{a}}\sin(2\delta\theta_{i})}
        }, \label{apx:eq:GammaEff} \\
        \Omega_{a}^{\text{eff}} &= \Omega_{a}\sqrt{1-\tfrac{\Gamma_{a}}{2\Omega_{a}}\sin(2\delta\theta_{i})}.  \label{apx:eq:LarmEff}
\end{align}
Equation~\eqref{apx:eq:QuadraturewQW_wCC} can be rewritten using Eqs.~\eqref{apx:eq:GammaEff} and \eqref{apx:eq:LarmEff} as the detected quadrature from an effective oscillator 
\begin{equation}
     \hat{Q}_{i}(\delta\theta_{i}) = g_\mathrm{VR}(\Omega) [
          \hat{p}_{i} 
          + \mathcal{K}_{a}^{\text{eff}}(\Omega) \hat{x}_{i} +\hat{N}_a^{\text{eff}}(\Omega)
          ],
    \label{apx:eq:b_s_eff}
\end{equation}
scaled by the factor 
\begin{equation}
    g_\mathrm{VR}(\Omega) \equiv 1-\mathcal{K}_{a}(\Omega, \Omega_{a},\Gamma_{a})\frac{\sin(2\delta\theta_{i})}{2},
    \label{apx:eq:VirtualRigidityGain}
\end{equation}
where the effective backaction contribution is governed by
\begin{equation}
    \mathcal{K}_{a}^{\text{eff}}(\Omega) \equiv \mathcal{K}_{a}(\Omega, \Omega_{a}^\mathrm{eff},\Gamma_{a}^\mathrm{eff}) =  \frac{\Gamma_{a}^{\text{eff}} \Omega_{a}^{\text{eff}}}{(\Omega_{a}^{\text{eff}})^2-\Omega^{2} -i \gamma^{}_{a}\Omega+ \gamma^{2}_{a}/4 },
    \label{apx:eq:EffBackAction}
\end{equation}
and the effective atomic noise associated with a finite decay is
\begin{equation}
    \hat{N}_a^{\text{eff}}(\Omega) \equiv \frac{\hat{N}_a(\Omega) \cos(\delta\theta_{i})}{g_\mathrm{VR}(\Omega)}.
\end{equation} 

\subsection{EPR light source}

Here we introduce the formal definitions for noise spectral densities employed in this work. We also specify the spectral densities characterizing the EPR source as well as extraneous sources of vacuum noise.

The symmetrized cross-correlation spectrum $S_{a,b}$ of two operators $a$ and $b$ is evaluated via
\begin{equation}\label{apx:eq:crossspectrum}
{S}_{a,b}(\Omega)\delta(\Omega-\Omega') = \frac{1}{2}\left<\hat{a}(\Omega)\hat{b}^{\dagger}(\Omega')+\hat{b}^{\dagger}(\Omega')\hat{a}(\Omega) \right>,
\end{equation}
which is generally complex-valued and obeys $S_{b,a}(\Omega)=S_{a,b}^*(\Omega)$; $\delta(\Omega-\Omega')$ is the Dirac delta function. 
The auto-correlation of an operator $\hat{a}$ amounts to the special case $\hat{a}=\hat{b}$ of Eq.~\eqref{apx:eq:crossspectrum},
\begin{equation}
{S}_{a}(\Omega) \equiv{S}_{a,a}(\Omega).
\end{equation}

According to the commutation relations  $[\hat{x}_{i}(t),\, \hat{p}_{i}(t')] = [\hat{x}_{s}(t),\, \hat{p}_{s}(t')]=i\delta(t-t')$, the symmetrised power spectral densities for vacuum fields are
\begin{gather}  
S_{SN}\equiv {S}_{v_x} = {S}_{v_p} = \frac{1}{2}, \\
{S}_{v_x,v_p} = 0.
\end{gather}
With this convention, the following spectral densities characterise the entangled outputs of the EPR source
\begin{gather}
   S_{{x}_{s}} = S_{{p}_{s}} = S_{{x}_{i}} =S_{{p}_{i}} = \frac{\cosh(2r)}{2} ,\\
   S_{{x}_{s},{x}_{i}} = - S_{{p}_{s} ,{p}_{i} }  = \frac{\sinh(2r)}{2};
    \label{apx:eq:crossCorellation}
\end{gather}
all other cross-spectral densities involving those operators are zero.

\subsection{Wiener filter}

As already mentioned in Methods, squeezing in the joint conditional signal $Q_{s|i}=q_{s}(\theta_s)+gQ_{i}(\theta_i)$ is optimized by varying $g(\Omega)$ using the Wiener filter theory. In the subsequent subsections we evaluate the squeezing performance on the basis of the model. 

For convenience, we restate the expression for the optimal filter and the resulting conditional squeezing here (also presented in Methods).
The optimal gain is given by 
\begin{equation}
    g(\Omega, \theta_{s}, \delta\theta_{i}) = - \frac{S_{q_{s},Q_{i}}(\Omega, \theta_{s}, \delta\theta_{i})}{S_{Q_{i}}(\Omega, \delta\theta_{i})},
\label{apx:eq:WienerGain}
\end{equation}
which leads to the power spectral density of the optimised signal 
\begin{equation}
\begin{aligned}
    S_{Q_{s|i}}(\Omega, \theta_{s}, \delta\theta_{i}) &= S_{q_{s}}(\Omega, \theta_{s}) - \frac{\left|S_{q_{s},Q_{i}}(\Omega, \theta_{s}, \delta\theta_{i})\right|^2}{S_{Q_{i}}(\Omega, \delta\theta_{i})}. \label{apx:eq:Opt_signal}
\end{aligned}
\end{equation}

\subsubsection{Atomic spin oscillator without virtual rigidity}
In the case $\delta\theta_i=0$, we are measuring the atomic spin oscillator without the quarter wave plate in the polarimetric homodyne detector. This configuration is described by
\begin{subequations}\label{apx:eq:IO_no-VR}
\begin{align}
    \hat{q}_{s}(\theta_{s}) &= \hat{x}_{s}\sin(\theta_{s}) +\hat{p}_{s}\cos(\theta_{s}),\\
    \hat{Q}_{i}(\delta\theta_{i}=0) &= \hat{p}_{i} + \mathcal{K}_{a} \hat{x}_{i}+ \hat{N}_a(\Omega).
\end{align}
\end{subequations}
By using Eqs.~\eqref{apx:eq:IO_no-VR} and noting that the only non-zero cross-correlations are given by Eqs.~\eqref{apx:eq:crossCorellation}, we find that
\begin{equation}
S_{q_{s},Q_{i}}(\Omega) = \frac{\sinh(2r)}{2}[-\cos(\theta_{s}) + \sin(\theta_{s}) \mathcal{K}_{a}^{*}(\Omega)].
\label{aps:eq:CrossCorr_simple_atom}
\end{equation}
The spectral density of the atomic spin oscillator signal is given by
\begin{equation}
\begin{aligned}
    S_{Q_{i}}(\Omega) &= S_{p_{i}} + \left|\mathcal{K}_{a}(\Omega)\right|^2  S_{{x}_{i}}  + {S}_{N}(\Omega)\\
    &=  \frac{\cosh(2r)}{2}\,(1 + \left|\mathcal{K}_{a}(\Omega)\right|^2) + {S}_{N}(\Omega),
\end{aligned}
\end{equation}
where ${S}_{N}$ is the spectrum of the uncorrelated noise, which will be described in Section~\ref{sc:RealisticModel}.

We can use Eq.~\eqref{apx:eq:WienerGain} to calculate the Wiener filter 
\begin{equation}
    g(\Omega,\theta_{s}, \delta\theta_{i}=0) =  \tanh(2r) \frac{
    \cos(\theta_{s}) - \sin(\theta_{s}) \mathcal{K}_{a}^{*}(\Omega)
    }{
    1+ \left|\mathcal{K}_{a}(\Omega)\right|^2 + 
    \tfrac{2{S}_{N}(\Omega)}{\cosh(2r)}
    },
\end{equation}
and Eq.~\eqref{apx:eq:Opt_signal} to estimate the conditional noise spectral density of the combined system as
\begin{equation}
    \frac{S_{Q_{s|i}}(\Omega, \theta_{s}, \delta\theta_{i}=0)}{S_{SN}} = \cosh(2r) -\tanh(2r)\sinh(2r) \frac{
    \left|
    \cos(\theta_{s}) - \sin(\theta_{s}) \mathcal{K}_{a}(\Omega)
    \right|^2}
    {
    1+ \left|\mathcal{K}_{a}(\Omega)\right|^2 
    +\tfrac{2{S}_{N}(\Omega) }{\cosh(2r)}
    },
    \label{eq:CondSQZ}
\end{equation}
where we have renormalised by the shot-noise spectral density of the idler arm $S_{SN}= 1/2$.
 
\subsubsection{Atomic spin ensemble with virtual rigidity}

The configuration $\delta\theta_i\neq 0$ corresponds to using the quarter wave plate in the polarimetric homodyne detection to manipulate the input-output relationship of the atomic spin oscillator.  The normalized signal and idler photocurrents are 
\begin{align}
    \hat{q}_{s}(\theta_{s}) &= \hat{x}_{s}\sin(\theta_{s}) +\hat{p}_{s}\cos(\theta_{s}),\\
    \hat{Q}_{i}(\delta \theta_{i}) &=  \hat{p}_{i} + \mathcal{K}_{a} \hat{a}_{i}^{\pi/2-\delta\theta_{i}}\cos(\delta\theta_{i}) + \hat{N}_a(\Omega)\cos(\delta\theta_{i}),
\end{align}
where $\hat{a}_{i}^{\pi/2-\delta\theta_{i}}= \hat{x}_{i}\cos(\delta\theta_{i}) +\hat{p}_{i}\sin(\delta\theta_{i})$. As we described in Section~\ref{sc:VRmodel}, we can rewrite the idler quadrature
\begin{equation}
     \hat{Q}_{i}(\delta \theta_{i}) = g_\mathrm{VR}(\Omega,\delta \theta_{i}) \left[
          \hat{p}_{i}  + \mathcal{K}_{a}^{\text{eff}} \hat{x}_{i} + \hat{N}_a^{\text{eff}} \right] = g_{VR}(\Omega,\delta\theta_{i})\hat{Q}_{i}^{\text{eff}},
\end{equation}
which corresponds to the response of an effective oscillator. 

The correlation between the photocurrents $S_{q_{s},Q_{i}}$ can be evaluated by noting that 
\begin{equation}
    \hat{q}_{s} \hat{Q}_{i}^{\dagger} = g_{VR}^{*} \left(\hat{q}_{s} (\hat{Q}_{i}^{\text{eff}})^{\dagger}\right).
\end{equation}

Based on this observation, we can re-adapt the result of Eq.~\eqref{aps:eq:CrossCorr_simple_atom} and calculate the cross-spectral density by replacing  $ \mathcal{K}_{a} \rightarrow \mathcal{K}_{a}^{\text{eff}}$
\begin{equation}
\begin{aligned}
S_{q_{s},Q_{i}} 
    &= g_\mathrm{VR}^{*}(\Omega) \frac{\sinh(2r)}{2}\,(-\cos(\theta_{s}) + \sin(\theta_{s}) [\mathcal{K}_{a}^{\text{eff}}(\Omega)]^{*}).
\end{aligned}
\end{equation}
Based on the same observation, the power spectral density of the atomic spin signal can be calculated as
\begin{equation}
    \begin{aligned}
    S_{Q_{i}}(\delta\theta_{i}) &= \left|g_\mathrm{VR}(\Omega)\right|^2 \left[\frac{\cosh(2r)}{2}\,(1 + \left|\mathcal{K}_{a}^{\text{eff}}(\Omega)\right|^2)
    +{S}_{N}^{\text{eff}}(\Omega)
    \right].
    \end{aligned}
\end{equation}
By combining the two equations, we can estimate the optimal gain
\begin{equation}
    \begin{aligned}
    g(\Omega,\theta_{s}, \delta\theta_{i}) 
    &= {g}_\mathrm{VR}^{-1}(\Omega)
    \,\tanh(2r)
    \,\frac{
    \cos(\theta_{s})
    - \sin(\theta_{s}) [\mathcal{K}_{a}^{\text{eff}}(\Omega)]^{*}
    }{
    1 + \left|\mathcal{K}_{a}^{\text{eff}}(\Omega)\right|^2 + \tfrac{2{S}_{N}^{\text{eff}}(\Omega)}{\cosh(2r)}
    }   ,  
    \end{aligned}
\end{equation}
and the power spectral density of the optimised signal in the units of shot noise is 
\begin{equation}
    \frac{S_{Q_{s|i}}(\Omega, \theta_{s}, \delta\theta_{i})}{S_{SN}}= \cosh(2r) -\tanh(2r)\sinh(2r) \frac{
    \left|
    \cos(\theta_{s}) - \sin(\theta_{s}) \mathcal{K}_{a}^{\text{eff}}(\Omega)
    \right|^2}
    {
    1+ \left|\mathcal{K}_{a}^{\text{eff}}(\Omega)\right|^2 
    + \tfrac{2{S}_{N}^{\text{eff}}(\Omega)}{\cosh(2r)}
    },
    \label{eq:CondSQZ_VR}
\end{equation}
where the normalisation factor ${g}_\mathrm{VR}(\Omega)$ is absorbed by the optimal gain $g(\Omega,\theta_{s}, \delta\theta_{i})$ and the dependency on $\delta \theta_{i}$ is contained in the effective quantities $\mathcal{K}_{a}^{\text{eff}}$ and ${S}_N^{\text{eff}}$.

\subsection{Realistic Model of the Atomic Spin Oscillator}\label{sc:RealisticModel}

\subsubsection{Spin oscillator thermal and broadband noise}
In the calculation presented until now, we have collected the uncorrelated noise contribution in the operator $\hat{N}_a(\Omega)$. Here, we provide the explicit expression and derive its noise spectral density $S_{N}(\Omega)$.  During the light-spin interaction, the spin oscillator experiences dissipation, which can be described using the quantum Langevin methods, as outlined in Refs.~\cite{Thomas2021, Jia2023}. The dissipation leads to the noise which has two main contributions: (1) The thermal noise, associated with a finite decay $\gamma_{a} > 0$, which is primarily due to the spontaneous emission of the probe photons, and (2) the broadband noise generated by imperfect coupling between the idler beam and the atomic spins moving within the vapour cell.
The thermal noise is given by $\mathcal{K}_{a}(\Omega) \hat{F}_{a}$, while the broadband noise is given by another frequency-dependent contribution $\mathcal{K}_{bb}(\Omega) \hat{F}_{bb}$. With those two contributions, the equation for the Stokes operator reads:
\begin{equation}\label{apx:eq:S_3-out}
\begin{aligned}
    \hat{\mathcal{S}}_{3}^\mathrm{out} &= \hat{\mathcal{S}}_{3}^{\text{in}} + \mathcal{K}_{a} \hat{\mathcal{S}}_{2}^{\text{in}} +{\mathcal{N}_{a}}\\ & = \hat{\mathcal{S}}_{3}^{\text{in}} + \mathcal{K}_{a} \hat{\mathcal{S}}_{2}^{\text{in}} + \left( \mathcal{K}_{th}(\Omega) \hat{F}_{a} + \mathcal{K}_{bb}(\Omega) \hat{F}_{bb} \right) \frac{\alpha_\textrm{pr}}{\sqrt{2}}.
\end{aligned}
\end{equation}
where
\begin{subequations}
\label{apx:eq:mathcalK-defs}
\begin{align}
\mathcal{K}_{th}(\Omega) &= \frac{\sqrt{2\gamma_{a}\Gamma_{a}}\Omega_{a}}{\Omega^{2}_{a} - \Omega^{2} - i \gamma_{a} \Omega + \gamma_{a}^{2} / 4},\\
    \mathcal{K}_{bb}(\Omega) &= \frac{\sqrt{2\gamma_{bb}\Gamma_{bb}}\Omega_{a}}{\Omega^{2}_{a} - \Omega^{2} - i \gamma_{bb} \Omega + \gamma_{bb}^{2} / 4},
\end{align}
\end{subequations}
where $\gamma_{a}$ and $\Gamma_{a}$ are the decay rate and the readout rate for the atomic narrowband response, and $\gamma_{bb}$ and $\Gamma_{bb}$ are the decay rate and the readout rate for the broadband dynamics. In our experiment, where $|\Omega_{a}|\gg\gamma_{a},\Gamma_{bb} $,  the thermal dynamics of the two transverse collective spin components contribute approximately equally to the output light, resulting in the factor of $\sqrt{2}$ in $\mathcal{K}_{th}(\Omega)$ and $\mathcal{K}_{bb}(\Omega)$. The spectrum of coupling to the thermal reservoirs, whose Langevin operators obey $[\hat{F}_{a}^{x}(t),\, \hat{F}_{a}^{p}(t')]= [\hat{F}_{bb}^{x}(t),\, \hat{F}_{bb}^{p}(t')] = i\delta(t-t')$, is characterised by the effective thermal occupation numbers $n_{th}$ and $n_{bb}$, \cite{Jia2023,Thomas2021}
\begin{equation}
S_{F_{a}} = n_{th}+\frac{1}{2},\quad
S_{F_{bb} }  = n_{bb}+\frac{1}{2}.
\label{apx:eq:mathcalK-spectrum-defs}
\end{equation}
The effective thermal occupation is linked to the internal spin state of the atomic ensemble. The purity of a coherent spin state, characterized by its spin polarization, can be quantified by using the
magneto-optical resonance method (MORS)\cite{julsgaard2003characterizing}. The experimental thermal occupation number is typically larger than the value estimated from the spin polarization. 
This is due to the influence of the classical probe intensity noise and environmental disturbances, as discussed in Ref.~\cite{Jia2023}. For purposes of fitting (via $\Gamma_{bb}$ and $\gamma_{bb}$) the broadband atomic noise profile $\propto\Gamma_{bb}(1/2+n_{bb})$ due to the contribution of $\hat{F}_{bb}$ in Eq.~\eqref{apx:eq:S_3-out}, we can take $n_{bb}=n_{th}$ as a matter of convention.

\subsubsection{Optical losses}

The losses in the idler arm have to be divided between the input efficiency $\eta_{i,\text{in}}$ and the output efficiency $\eta_{i,\text{out}}$, which correspond to the optical losses acquired before and after the interaction with the atomic spin oscillator, respectively.
The losses modify the input Stokes quadratures as
\begin{align}
    \hat{\mathcal{S}}_{2}^{\text{in}}\sqrt{2}/\alpha_\textrm{pr} &= \sqrt{\eta_{i,\text{in}}} \hat{q}_{\text{i}}(\phi_{i})+ \sqrt{1-\eta_{i,\text{in}}} \hat{v}_{i,\text{in}}(\phi_{i}), \\
    \hat{\mathcal{S}}_{3}^{\text{in}}\sqrt{2}/\alpha_\textrm{pr} &= \sqrt{\eta_{i,\text{in}}} \hat{q}_{\text{i}}(\phi_{i}+\pi/2) + \sqrt{1-\eta_{i,\text{in}}} \hat{v}_{i,\text{in}}(\phi_{i}+\pi/2),
\end{align}
where, $\hat{v}_{i,\text{in}}(\phi_{i})$ and $\hat{v}_{i,\text{in}}(\phi_{i}+\pi/2)$ are the quadratures of the vacuum entering before the interaction with the atomic spin oscillator. The output losses correspond to injection of another vacuum state $\hat{v}_{i,\text{out}}^{0}$ which lead to measured Stokes quadrature operator
\begin{equation}
    \hat{\mathcal{R}_{i}(\delta\theta_{i})}   =  \sqrt{\eta_{i,\text{out}}} \left(\hat{\mathcal{S}}_{2}^\text{out} \cos(\delta\theta_{i}) + \hat{\mathcal{S}}_{3}^\text{out} \sin(\delta\theta_{i})\right) + \alpha_\textrm{pr}/\sqrt{2} \sqrt{1-\eta_{i,\text{out}}} \hat{v}_{i,\text{out}}^{0}  = \alpha_\textrm{pr}/\sqrt{2}\, \hat{Q}_{i}(\delta\theta_{i}).
\end{equation}

The final expression for the photocurrent is quite crowded, and for the sake of simplicity, we will derive the results first without the quarter wave plate, $\delta\theta_{i}=0$, and then in the case when it is included, $\delta\theta_{i}\neq 0$.

\subsubsection{Realistic model without virtual rigidity}

In the absence of virtual rigidity effect, $\delta\theta_{i}=0$, the input idler phase $\phi_{i}=0$, and the idler photocurrent is
\begin{multline}
     \hat{Q}_{i}(\delta\theta_{i}=0) 
     = \sqrt{\eta_{i,\text{out}}} \Bigg[
     \sqrt{\eta_{i,\text{in}}} \left( \hat{p}_{i}  - \mathcal{K}_{a} \hat{x}_{i} \right) 
     +\sqrt{1-\eta_{i,\text{in}}} \left( \hat{v}_{i,\text{in}}^{0} - \mathcal{K}_{a} \hat{v}_{i,\text{in}}^{\pi/2} \right) \\
           + \mathcal{K}_{th}(\Omega) \hat{F}_{a} 
          + \mathcal{K}_{bb}(\Omega) \hat{F}_{bb} + \frac{\sqrt{1-\eta_{i,\text{out}}}}{\sqrt{\eta_{i,\text{out}}}} \hat{v}_{i,\text{out}}^{0}\Bigg],
\label{apx:eq:photocurrentIdler_real}
\end{multline}
where we can identify four new contributions to the final photocurrent:
\begin{equation}
    \hat{\Sigma}_{i,\text{in}} = \sqrt{1-\eta_{i,\text{in}}} \left( \hat{v}_{i,\text{in}}^{0} - \mathcal{K}_{a} \hat{v}_{i,\text{in}}^{\pi/2} \right),
\end{equation}
is the response of the atomic spin oscillator to the vacuum appearing due to the sub-optimal coupling; 
\begin{equation}
    \hat{\Sigma}_{\text{th}} = \mathcal{K}_{th}(\Omega) \hat{F}_{a},
\end{equation}
is the thermal noise; 
\begin{equation}
    \hat{\Sigma}_{\text{bb}}= \mathcal{K}_{bb}(\Omega) \hat{F}_{bb},
\end{equation}
 is the broadband noise; and 
\begin{equation}
  \hat{\Sigma}_{i,\text{out}} = \hat{v}_{i,\text{out}}^{0}\,\sqrt{1-\eta_{i,\text{out}}}/\sqrt{\eta_{i,\text{out}}},    
\end{equation}
is the effect of imperfect readout efficiency.

The new terms will affect the power spectral density of the idler signal by adding new contributions  
\begin{subequations}\label{apx:eq:Lambda-defs}
\begin{align}
     \Lambda_{\text{in}}(\Omega) &\equiv S_{\Sigma_{i,\text{in}}} (\Omega)  = \frac{1-\eta_{i,\text{in}}}{2} \left(1+ \left|\mathcal{K}_{a}(\Omega)\right|^2\right),\\
     \Lambda_{\text{out}}(\Omega)&\equiv S_{\Sigma_{i,\text{out}}}(\Omega)  = \frac{1-\eta_{i,\text{out}}}{2\eta_{i,\text{out}}},\\
     {S}_{\text{th}}(\Omega)&\equiv S_{\Sigma_{\text{th}}}(\Omega)     = \left|\mathcal{K}_{th} (\Omega)\right|^2  (1/2+n_{\text{th}}),\\
     {S}_{\text{bb}}(\Omega)&\equiv S_{\Sigma_{\text{bb}}}(\Omega)     = \left|\mathcal{K}_{bb} (\Omega)\right|^2  (1/2+n_{\text{bb}}),     
\end{align}
\end{subequations}
which add up to
\begin{equation}
    S_{Q_{i}} = \eta_{i,\text{out}} \Bigg[ \frac{\cosh(2r)\eta_{i,\text{in}}}{2} \left(1+ \left|\mathcal{K}_{a}\right|^2\right) + \frac{1-\eta_{i,\text{in}}}{2} \left(1+ \left|\mathcal{K}_{a}\right|^2\right) 
+  S_{\Sigma_{\text{th}}}
+  S_{\Sigma_{\text{bb}}}
+  \frac{1-\eta_{i,\text{out}}}{2 \eta_{i,\text{out}}} \Bigg].
\end{equation}
The cross-correlation between the photocurrents is affected only by the losses on the idler and the signal, and it is reduced as 
\begin{equation}
    S_{q_{s},Q_{i}} = - \sqrt{\eta_{s}\eta_{i,\text{out}}\eta_{i,\text{in}}} \,
     \frac{\sinh(2r)}{2}\,(\cos(\theta_{s}) - \sin(\theta_{s}) [\mathcal{K}_{a}(\Omega)]^{*}),
\end{equation}
and the spectrum of the signal photocurrent is
\begin{equation}
    S_{{q}_{s}}= \eta_{s}\,\frac{\cosh(2r)}{2} + \frac{1-\eta_{s}}{2},
\end{equation}

The power spectral density of the optimised signal is evaluated using Eq.~\eqref{apx:eq:Opt_signal} as
\begin{equation}
    \frac{S_{Q_{s|i}}(\Omega, \theta_{s}, \delta\theta_{i}=0)}{S_{SN}} = 1-\eta_{s} + \frac{\eta_{s}}{\cosh(2r)} 
\Bigg[
\cosh^2(2r) - 
   \frac{
    \sinh^2(2r)
   \left|\cos(\theta_{s}) - \sin(\theta_{s}) \mathcal{K}_{\text{a}} \right|^2
    }{
    1+ \left|\mathcal{K}_{\text{a}}\right|^2
    +2\tfrac{\Lambda_{\text{in}}+\Lambda_{\text{out}} + 
        {S}_{\text{th}} 
         +{S}_{\text{bb}} 
    }{\eta_{i,\text{in}} \cosh(2r)}
    }
\Bigg]. \label{apx:eq:OptimalSpectrum}
\end{equation}
in terms of the quantities~\eqref{apx:eq:Lambda-defs}, highlighting the contributions of different imperfections (dependencies on $\Omega$ are omitted for brevity).

\subsubsection{Realistic model with virtual rigidity}

Even in the realistic model, we can use the virtual rigidity renormalisation, Eq.~\eqref{apx:eq:VirtualRigidityGain}, to reinterpret the effect of the quarter waveplate as a shift of the resonance frequency, Eq.~\eqref{apx:eq:LarmEff} and rescaling of the readout rate, Eq.~\eqref{apx:eq:GammaEff} on the backaction coefficient $\mathcal{K}_{a}$.  Here, we discuss the effect of the virtual rigidity renormalisation on the thermal noise, the broadband noise and the optical losses. 
In the virtual rigidity configuration, the idler photocurrent is proportional to:
\begin{multline}
    \hat{Q}_{i}(\delta\theta_{i}) 
     = \sqrt{\eta_{i,\text{out}}} g_\mathrm{VR}(\Omega) \Bigg[
     \sqrt{\eta_{i,\text{in}}} \left( \hat{p}_{i}  - \mathcal{K}_{a}^{\text{eff}}(\Omega) \hat{x}_{i} \right) 
     +\sqrt{1-\eta_{i,\text{in}}} \left( \hat{v}_{i,\text{in}}^{0} - \mathcal{K}_{a}^{\text{eff}}(\Omega) \hat{v}_{i,\text{in}}^{\pi/2} \right) \\
           + \mathcal{K}_{th}^{\text{eff}}(\Omega) \hat{F}_{a} 
          + \mathcal{K}_{bb}^{\text{eff}}(\Omega) \hat{F}_{bb} + \frac{\sqrt{1-\eta_{i,\text{out}}} }{\sqrt{\eta_{i,\text{out}}}g_\mathrm{VR}(\Omega)}\hat{v}_{i,\text{out}}^{0} \Bigg],
\label{apx:eq:photocurrentIdler_real_VR}    
\end{multline}
where $g_\mathrm{VR}(\Omega)$ is the virtual rigidity gain defined in Eq.~\eqref{apx:eq:VirtualRigidityGain}, $\mathcal{K}_{a}^{\text{eff}}$ is the effective atomic transfer function defined in Eq.~\eqref{apx:eq:EffBackAction}, and $\mathcal{K}_{th}^{\text{eff}}(\Omega)$ and $\mathcal{K}_{bb}^{\text{eff}}(\Omega)$ are the effective transfer functions for the thermal and broadband noise,
\begin{align}
    \mathcal{K}_{th}^\mathrm{eff} (\Omega)&\equiv  \frac{\cos(\delta\theta_{i})}{g_\mathrm{VR}(\Omega)}  \mathcal{K}_{th}(\Omega),  \\
    \mathcal{K}_{bb}^\mathrm{eff}(\Omega) &\equiv \frac{\cos(\delta\theta_{i})}{g_\mathrm{VR}(\Omega)}  \mathcal{K}_{bb}(\Omega).
\end{align}

The Eqs.~\eqref{apx:eq:photocurrentIdler_real_VR} and \eqref{apx:eq:photocurrentIdler_real} are similar, and are related by the following replacements
\begin{align}
    \mathcal{K}_{th} &\rightarrow \mathcal{K}_{th}^{\text{eff}}, \\
    \mathcal{K}_{bb} &\rightarrow \mathcal{K}_{bb}^{\text{eff}}, \\
    \frac{\sqrt{1-\eta_{i,\text{out}}}}{\sqrt{\eta_{i,\text{out}}}} &\rightarrow \frac{\sqrt{1-\eta_{i,\text{out}}}}{\sqrt{\eta_{i,\text{out}}}g_\mathrm{VR}(\Omega)},
\end{align}
which lead to the effective noise contribution factors [cf.\ Eqs.~\eqref{apx:eq:Lambda-defs}]
\begin{subequations}
\begin{align}
    \Lambda^{\text{eff}}_{\text{in}}     (\Omega)&= \frac{1-\eta_{i,\text{in}}}{2}(1+ \left|\mathcal{K}^{\text{eff}}_{a}(\Omega)\right|^2), \\
    \Lambda^{\text{eff}}_{\text{out}}   (\Omega) &= \frac{1-\eta_{i,\text{out}}}{2\eta_{i,\text{out}}} \frac{1}{|g_\mathrm{VR}(\Omega)|^2},  \\ 
    {S}^{\text{eff}}_{\text{th}}(\Omega) &= \left|\frac{\cos(\delta\theta_{i})\mathcal{K}_{th} (\Omega)}{g_\mathrm{VR}(\Omega)}\right|^2 (1/2+n_{\text{th}}), \\ 
    {S}^{\text{eff}}_{\text{bb}}(\Omega) &= \left|\frac{\cos(\delta\theta_{i})\mathcal{K}_{bb} (\Omega)}{g_\mathrm{VR}(\Omega)}\right|^2 (1/2+n_{\text{bb}}).
\end{align}
\end{subequations}
These effective noise contributions are inserted in Eq.~\eqref{apx:eq:OptimalSpectrum}
\begin{equation}
    \frac{S_{{Q}_{s|i}}(\Omega, \theta_{s}, \delta\theta_{i})}{S_{SN}} = 1-\eta_{s} + \frac{\eta_{s}}{\cosh(2r)} 
\left[
\cosh^2(2r) -
   \frac{
    \sinh^2(2r)
   \left|\cos(\theta_{s}) - \sin(\theta_{s}) \mathcal{K}_{a}^{\text{eff}} \right|^2
    }{
    1+ \left|\mathcal{K}_{a}^{\text{eff}}\right|^2
    +2\tfrac{ \Lambda^{\text{eff}}_{\text{in}} +\Lambda^{\text{eff}}_{\text{out}} + 
        {S}^{\text{eff}}_{\text{th}}
         +{S}^{\text{eff}}_{\text{bb}}
     }{\eta_{i,\text{in}} \cosh(2r)}
    }
\right], 
\end{equation}
where, on the right-hand side, the dependency on $\delta\theta_i$ is contained in the effective quantities (`eff') and the dependencies on $\Omega$ are omitted for brevity.

\section{Data processing}

\subsection{Wiener filter implementation}

In the previous section, we developed the analytical model to predict the noise spectra for joint continuous measurements in the signal and idler channels using the optimal Wiener gain to achieve the maximum conditional squeezing. This frequency-dependent optimal weighting can be determined experimentally through the cross-correlations between two entangled channels, as discussed in Refs.~\cite{Ma2017,Danilishin2019,gould2021optimal}. Here we implement this procedure and present the post-processing method which allows us to infer the optimal conditional squeezing directly using the cross-correlation between the recorded signal and idler time series. This can be contrasted with Wiener-filter protocols for inferring the state of a system coupled to light, which requires a model for the system dynamics and input-output relations \cite{wieczorek2015optimal,rossi2019observing,meng2022measurement,Thomas2021}. Nevertheless, the predicted conditional squeezing---based solely on the cross-correlation between the signal and idler channels---provides a cross-validation method to test the agreement of our theoretical model with the calibrated parameters as discussed in the main text.

\begin{figure}[ht!]
\centering
\includegraphics[width=0.9\textwidth]{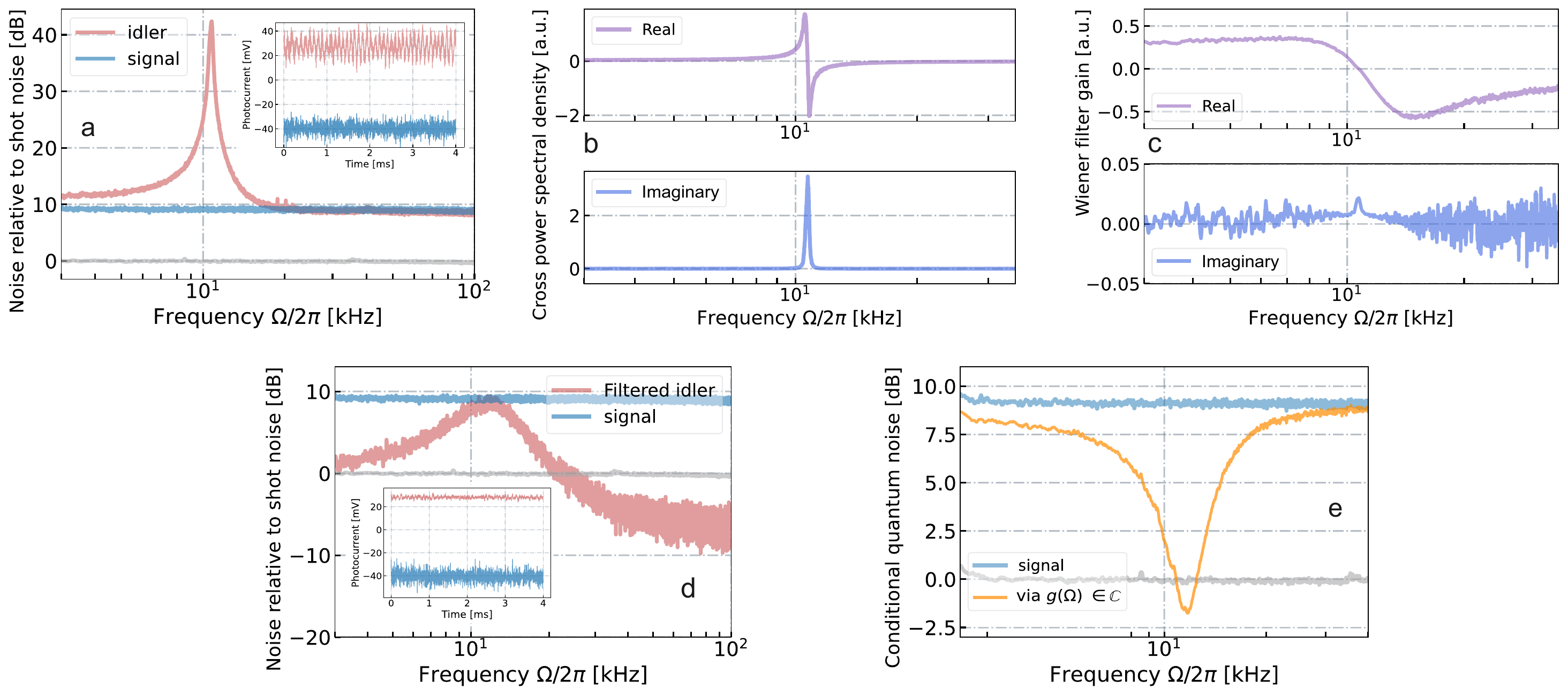}
\caption{\textbf{Conditional squeezing by Wiener filtering based on the measured cross-correlation spectrum of entangled channels.} The signal analysis is performed with the signal measured at near amplitude quadrature ($\theta_{s} =  80^{\circ}$). The spin oscillator is operated in the positive-mass configuration and the idler field is detected at phase quadrature ($\delta\theta_{i} =  0^{\circ}$). (a) Recorded photocurrents of the signal (blue curve) and idler (red curve) fields are presented over a 4\,ms window out of a total 60\,s measurement, along with their power spectral densities estimated via Welch's method. (b) The time series of the signal and idler are used to calculate the complex cross-power spectral density using Welch's method, The real (purple curve) and imaginary (blue curve) components are shown. (c) The optimal complex Wiener filter gain $g(\Omega)$ is computed using the estimated cross-power spectral density and the idler's autocorrelation function. (d) The experimentally determined optimal filter gain is applied to the idler data, enabling the estimation of the filtered idler signal (shown by the red trace). (e) The original signal time series and the filtered idler series are combined to estimate the optimal conditional squeezing level (orange curve). 
}
\label{f:data analysis}
\end{figure}

The recorded photocurrents from both signal and idler detectors are saved as time series data, as shown in Fig.~\ref{f:data analysis}(a). The corresponding power spectral densities (PSD) are calculated using Welch's method. Additionally, we estimate the complex cross power spectral density (CSD) between the idler and signal time series under the same PSD parameters. This estimated CSD, together with the previously computed idler power spectrum density, is used to design the Wiener filter gain $g(\Omega) = - S_{q_{s},Q_{i}}(\Omega)/S_{Q_{i}}(\Omega)$. We then apply this filter gain by convolving it with the idler beam time series, thereby optimally suppressing uncorrelated noise and compensating for the frequency-dependent phase lag arising from the damping of the atomic spin oscillator (highlighted by the imaginary components in royal blue trace). The 
resulting filtered idler time series represented by the light red curve in Fig.~\ref{f:data analysis}(d)] is subsequently combined with the original signal (blue curve) to infer the minimal conditional squeezing level. As shown in Fig.~\ref{f:data analysis}(e), the conditional spectrum (the orange curve) is significantly suppressed compared to the original signal's PSD (the blue curve).

Therefore, beyond the broadband quantum noise reduction achievable when both the measurement rate and complex susceptibility of the spin oscillator are matched with the target sensors, the reported conditional quantum noise reduction of the signal field at different detection phases---effectively treated as a sensor with purely real frequency response---opens up the potential for achieving broadband quantum noise reduction in other systems, even those without intrinsic damping (such as free-mass gravitational wave detectors).

\FloatBarrier

\begin{figure}[ht!]
\centering
\includegraphics[width=0.98\textwidth]{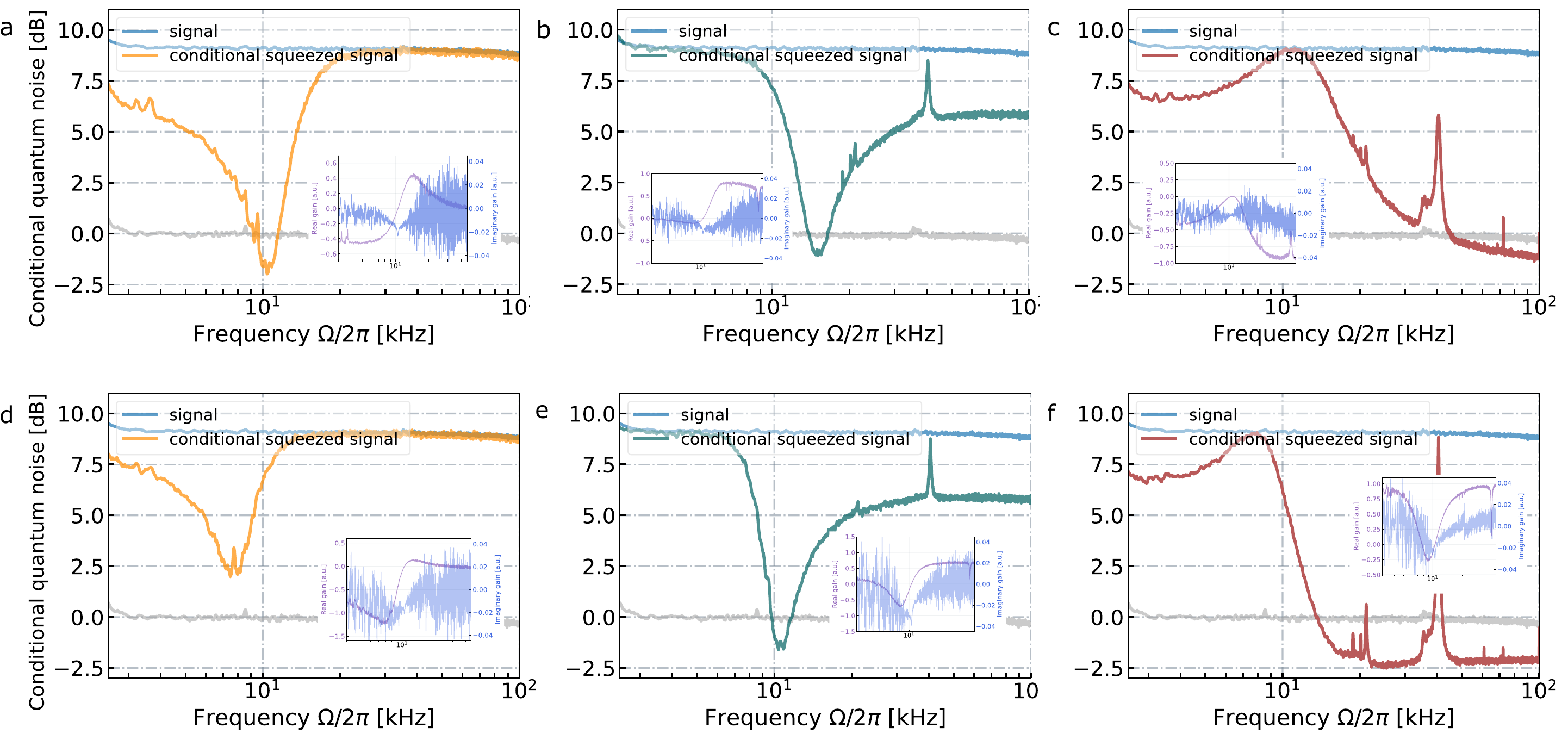}
\caption{\textbf{Conditional squeezing with and without virtual rigidity shift for different signal detection phases $\theta_s$.} The atomic spin oscillator is operated in the negative-mass configuration. (a-c) Conditional squeezing level of the signal beam is estimated using the optimal Wiener gain based on the idler beam measured in the phase quadrature. The signal beam is detected at the phase of $90^{\circ}$ (orange), $140^{\circ}$ (teal), $180^{\circ}$ (red), respectively, each with their corresponding complex Wiener gains [represented by their real (purple) and imaginary (blue) components]. (d-f) Introducing a quarter waveplate enables a virtual frequency downshift, altering the optimal Wiener gain and shifting the frequency corresponding to the minimal conditional squeezing level for each signal detection phase. The minimal quantum noise level remains nearly unchanged for the configuration without virtual rigidity. However, with the virtual rigidity shift applied, the minimal conditional squeezing level exhibits dependence on the Fourier frequency.}
\label{f:data analysis with different angle}
\end{figure}

In Fig.~\ref{f:data analysis with different angle}(a,b,c), we present the minimal conditional, frequency-dependent squeezing for three different signal homodyne phases (90$^{\circ}$,140$^{\circ}$,180$^{\circ}$), along with their corresponding complex Wiener gains. Furthermore, as illustrated in Fig.~\ref{f:data analysis with different angle}(d,e,f), activating the virtual rigidity in the idler arm we observe a clear downshift in the frequency associated with the minimal squeezing level compared to the configurations without virtual rigidity for all three phases presented in the Figure. The value of the frequency downshift varies for different signal readout angles, which can be attributed to a reduced effective atomic readout rate. The effective downshift is accompanied by asymmetric quantum noise reduction. Specifically, above the original Larmor frequency, we observe the enhanced $\approx -2.5$\,dB conditional squeezing [red curve in Fig.~\ref{f:data analysis with different angle}(f) compared to Fig.~\ref{f:data analysis with different angle}(c)].

\FloatBarrier
\subsection{Expected broadband quantum noise reduction with reduced imperfections}

Potential enhancement of the hybrid conditional frequency-dependent squeezing can be achieved by reducing losses and eliminating extra noise. Extended Data Fig.~\ref{fig:conditional squeezing with improved system} presents the minimal noise spectrum obtained from joint measurements optimized across various signal homodyne phases (red data points). Our theoretical model with calibrated parameters from the current experiment fits the experimental data well. Using this model, we can predict the performance improvements achievable by reducing various imperfections. 
A reduction in the effective thermal occupation number from its initial value of $n_{th}\approx 3.5$ by a factor of 3 primarily reduces quantum noise around the Larmor frequency (purple curve). This improvement can be realized through enhanced optical pumping and further suppression of classical noise sources, including those from the probe laser and acoustic magnetic noise from the environment. Additionally, decreasing the atomic broadband noise readout rate by a factor of 6, as can be achieved by increasing the filling factor of the top-hat beam, reduces the quantum noise both above and below the Larmor frequency (green curve). Combining these improvements, we anticipate achieving the 3\,dB conditional frequency-dependent squeezing achieved with the atomic spin oscillator limited only by the degree of squeezing of light.

\FloatBarrier

\section{Comparison of the spin system and a filter cavity}

Consider a Fabry-Pérot filter cavity with a distance $L_f$ between the two mirrors. In the low-intracavity-loss approximation, the quadrature phase shift of a detuned filter cavity can be approximated by \cite{evans2013realistic,kwee2014decoherence,McCuller2020}

\begin{equation}
\Phi_{f}(\Omega)\approx\arctan\left[\frac{2 \delta_f \gamma_f}{\gamma_f^2-\delta_f^2+\Omega^2}\right],
\label{eq:phase_fc}
\end{equation}
where $\Omega$ is the sideband frequency, $\delta_f$ is the filter cavity detuning from the field carrier frequency and $\gamma_f$ is the filter cavity bandwidth.  

As an example, we consider a filter cavity with finesse $\mathcal{F} = 6000$, a value similar to that reported in Ref.~\cite{Ganapathy2023}. To compare the transformation of the squeezed state by a filter cavity and by our system, we fit the readout angle of the conditional frequency-dependent noise spectra shown in Fig.~\ref{fig:EntAtomsHybrid10kHzPosNegMasVR} of the main text with Eq.~\eqref{eq:phase_fc} for different spin conditions to extract the parameters $\delta_f$ and $\gamma_f$, see Fig.~\ref{fig:fit_phase_fc} below. 

We use the estimated bandwidth $\gamma_f$ to calculate the cavity length as $L_f \approx c/(2\gamma_f\mathcal{F})$. For the case of Fig.~\ref{fig:EntAtomsHybrid10kHzPosNegMasVR}(d) of the main text, the fit (Fig.~\ref{fig:fit_phase_fc}, left panel) yields $\gamma_f/2\pi= 11.5$\,kHz, $\delta_f/2\pi= 4.7$\,kHz corresponding to $L_f\approx 5$\,m. For the spin system with virtual rigidity shown in Fig.~\ref{fig:EntAtomsHybrid10kHzPosNegMasVR}(f) of the main text, the filter cavity with equivalent performance requires $\gamma_f/2\pi= 8.1$\,kHz and $\delta_f/2\pi = 2.7$\,kHz resulting in the length $L_f\approx 10$\,m, as illustrated with the fit in the right panel of Fig.~\ref{fig:fit_phase_fc}. These values closely match the extrapolated effective Larmor frequency $\lvert\Omega_{a}^{\text{eff}}\rvert/2\pi= 7.9$\,kHz and half effective readout rate $\Gamma_{a}^{\text{eff}}/4\pi = 2. 8$\,kHz.

\begin{figure}[ht!]
\centering
\includegraphics[width=0.99\textwidth]{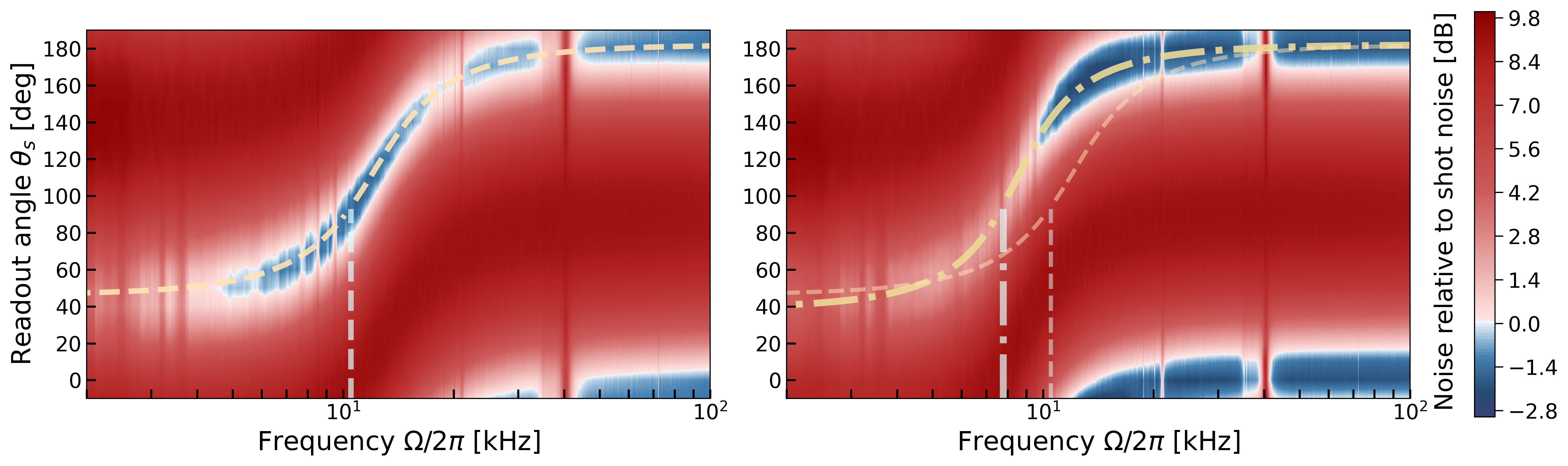}
\caption{\textbf{Comparison of the frequency-dependent squeezing generated by our spin system and the model for an analogous filter cavity, Eq.~\eqref{eq:phase_fc}.} See explanations in the text.} 
\label{fig:fit_phase_fc}
\end{figure}

\bibliographystyle{naturemag_format_title}
\bibliography{references_SI.bib}